\documentclass[a4paper,10pt]{article}

\usepackage{amssymb}

\usepackage{algorithmic}
\usepackage{algorithm}
\usepackage{appendix}
\usepackage[english]{babel}
\usepackage{cancel}
\usepackage{color}
\usepackage{graphicx}
\usepackage{hyperref}
\usepackage{ifthen}
\usepackage[utf8x]{inputenc}
\usepackage{subfig}
\usepackage{xfrac}

\newboolean{showComments}
\setboolean{showComments}{true}
\newcommand{\todocmd}[1]{{\bf\textcolor{red}{#1}}}
\newcommand{\todo}[1]{\ifthenelse {\boolean{showComments}} {\todocmd{#1}} {}}
\newcommand{\todof}[1]{\ifthenelse {\boolean{showComments}} {\footnote{\todocmd{#1}}} {}}

\renewcommand{\v}[1]{\ensuremath{\mathbf{#1}}} 

\newcommand{\abs}[1]{\left| #1 \right|} 
\newcommand{\avg}[1]{\left< #1 \right>} 
\newcommand{\pd}[2]{\frac{\partial #1}{\partial #2}}

\newcommand{\braket}[2]{\left< #1 \vphantom{#2} \right | \left. #2 \vphantom{#1} \right>} 
\let\baraccent=\= 
\renewcommand{\=}[1]{\stackrel{#1}{=}} 
\providecommand{\e}[1]{\ensuremath{\times10^{#1}}}
\newcommand{\prog}[0]{CLTDSE }

\usepackage{listings}

\newcounter{bla}

\usepackage{authblk}
\begin{document}




\title{A GPGPU based program to solve the TDSE in intense laser fields through the finite difference approach.}

\author{Cathal \'O Broin}
\author{L.A.A Nikolopoulos}

\affil{School of Physical Sciences, Dublin City University and National Centre for Plasma Science and Technology,  Ireland}
 \maketitle
 
\begin{abstract}
We present a \emph{General-purpose computing on graphics processing units} (GPGPU) based computational program and framework for the electronic dynamics of atomic systems under intense laser fields. We present our results using the case of hydrogen, however the code is trivially extensible to tackle problems within the single-active electron (SAE) approximation.  Building on our previous work, we introduce the first available GPGPU based implementation of the Taylor, Runge-Kutta and Lanczos based methods created with strong field \textit{ab-initio} simulations specifically in mind; \prog. The code makes use of finite difference methods and the OpenCL framework for GPU acceleration. The specific example system used is the classic test system; Hydrogen. After introducing the standard theory, and specific quantities which are calculated, the code, including installation and usage, is discussed in-depth. This is followed by some examples and a short benchmark between an 8 hardware thread (i.e logical core) Intel Xeon CPU and an AMD 6970 GPU, where the parallel algorithm runs 10 times faster on the GPU than the CPU.
\end{abstract}

\newpage

\section{Introduction}
Currently, the exploration of fundamental processes that atomic and molecular systems undergo when under the action of intense laser fields is a major research area. Such processes are studied, through experiment, using infrared wavelengths \cite{Posthumus2004} or at short wavelengths by free electron lasers (FELs) \cite{WabnitzEtAl2005}. Modelling \textit{Ab initio} the time-dependent dynamics of multielectron systems under intense time dependent fields to high accuracy is a computationally demanding task. Nevertheless, many \textit{ab initio} methods, which are outside the perturbation regime, have been developed to solve the TDSE with few electrons, such as ~\cite{DundasEtAl2000, PengEtAl2003, BarmakiEtAl2003, AwasthiEtAl2005, PalaciosEtAl2005, PalaciosEtAl2006, Sanz-VicarioEtAl2006, CaillatEtAl2005}.

Computational tractability, combined with few approximations, and accuracy are the most desirable properties when studying multielectron systems, \textit{ab-initio}, under intense ultrashort electromagnetic fields. Many approaches have been developed in atomic and molecular physics, most notably, Time-Dependent Hartree-Fock (TDHF) \cite{Kulander1987} and Time-Dependent Density Functional Theory (TDDFT) \cite{GiovanniEtAl2002}.

An alternative approach, with a long and considered history, is the adoption of the single-active-electron (SAE) approximation for atoms \cite{Kulander1987}. Models based on the SAE approximation, which reduces the dimensionality of the problem by freezing the tightly bound inner electrons, have a proven track record in cases where multiple electron excitations are not important and where single electron effects, such as above threshold ionisation (ATI) and high harmonic generation (HHG), dominate.

In the present work, we present a parallel finite-difference code (\prog) which can run on arbitrary computing devices (CPU, GPU, FPGAs etc), provided OpenCL support is available, and which works within the SAE context. Although we present results on hydrogen, it is a simple matter to employ an effective single-electron potential and treat the code within the SAE approximation.

\prog is a free software, C99 and OpenCL C based, GPGPU package which implements the action of the time evolution operator on the state of a system described by the Time Dependent Schr\"odinger Wave Equation (TDSE). The code implements three types of integration methods; embedded and non-embedded Runge-Kutta methods, the arbitrary order Taylor propagator, and Lanczos based propagation. The Lanczos propagator does not explicitly prescribe a specific integration method, but rather it facilitates the integration process by changing the system of equations one needs to integrate to a smaller one. In the current case, the Lanczos method is combined with a Taylor propagator.

In our previous paper \cite{OBroinNikolopoulos2012} we introduced the topic of OpenCL based GPU acceleration of the Time-Dependent Schr\"odinger Equation (TDSE) within the context of strong field physics. The suitability of performing \textit{ab initio} calculations for GPU acceleration was focused on the basis case where a significant performance improvement was achieved against a standard serial implementation for the case of the Runge-Kutta methods and an arbitrary order Taylor series expansion. To contrast with the previous case, we add the Lanczos method to the methods implemented, and the discussion and code released is centered around finite difference methods. In the context of ab initio calculations, this invovles very sparse matrices.

GPU parallelisation efforts are not in accelerating the integration method itself; the Runge-Kutta, Lanczos and Taylor methods are serial, and the calculation of each derivative is done sequentially. That is, both methods require multiple derivatives to be calculated, where each new derivative calculation requires information from the previous derivative. Rather than using a parallel integration algorithm, the matrix vector calculation that is performed at each time step is, itself, parallelised. GPUs are the natural partner to this form of embarrassingly parallel linear algebra acceleration.

The successful ab-initio time integration of the TDSE is seen as a very computationally demanding task and numerous different methods exist for the treatment of non-perturbative systems \cite{GuanEtAl2009, Muller1999, KulanderEtAl1992, DundasEtAlEurPhysJ2003, Moore2EtAl2011, NikolopoulosEtAl2008, Peng2006}. Many methods use, incorporate or are derivatives of the finite difference approach.

QPROP \cite{BauerKoval2006} is a C++ library which helps to hide the complexity of split-operator propagation for TDSE and TDDFT systems. That is, it exposes propagator functionality, but the specific application is built by the end user, albeit at greatly reduced complexity. QPROP is under a non-commercial use license. Non-commercial licenses do not meet the open source definition of the Open Source Initiative\textsuperscript{\textregistered} or the free software requirements of the Free Software Foundation\textsuperscript{\textregistered}. PyProp \cite{Birkeland2009} is a free software, grid based, TDSE propagator written in a mixture of C++, Fortran and Python. It provides a finite difference grid based method, amongst other grid types, as well as providing Krylov subspace based propagation. PyProp is under a copyleft license, but does not support GPUs directly, and public development appears to have stopped.

The HELIUM code, is a Fortran based code for 2 electron atomic systems. In terms of its code, it takes a complex, multi-layered approach, appearing to have 5 relatively independent logical layers, and thus at least 4 separate levels of function wrapping \cite{SmythEtAl1998}. The HELIUM code does not appear to be freely available. The approach taken in the current case for \prog has been to tie the propagator functionality to OpenCL; being too generic in operation has its own performance penalty. In principle the system could be quite easily adapted to the language CUDA and much of the propagator code was originally written in pure C99 for the basis case. This has been spun off now, as a stand alone separate program.

ALTDSE (Arnoldi-Lanczos TDSE) \cite{GuanEtAl2009} is another non-commercial licensed package. While it is a basis method, it is noteworthy because it uses a Krylov method; the more general Arnoldi method, written in Fortran 95.  2dhf is a recently updated \cite{Kobus2013} Fortran 77 electronic structure program. It makes use of a high order finite difference scheme combined with a mixed Self consistent fields (SCF) and coloured successive over-relaxation (SOR, a variant of standard Gauss-Seidel iteration) approach.

QnDynCUDA is a C++ class available under the CPC non-profit use licence agreement, which provides functionality to solve the TDSE using a method which relies on FFT and Chebyshev polynomials \cite{QnDynCUDA}. The code has been extended to take advantage of CUDA. CUDA is limited only to Nvidia discrete GPU technology. Thus the future of any CUDA code is dependent on future fortunes of Nvidia, and its future within high performance computing.

In the last decade high performance computational infrastructure has moved away from custom fully integrated systems towards distributed computing models based on highly interconnected commodity machines. At present, these distributed systems are being augmented with various forms of accelerators. Accelerators focus on optimising a specific class of problem. The strength of GPUs is in optimising throughput focused and computationally heavy problems. Currently GPGPU usage is mostly of discrete cards, but with the current ongoing convergence between CPUs and GPUs, it is expected that improved GPGPU performance with double precision will be available on integrated chips. Two benefits of this integration will be the lower latency and higher bandwidth between the GPU and CPU.

GPUs use a Single-Instruction Multiple-Data (SIMD) layout, or very long instruction words (VLIW), or a combination of the two. In a SIMD, a processor takes one instruction with multiple data arguments and executes the instructions on multiple processing elements. A VLIW is similar in operation except that each processing element can be fed a different instruction. The AMD FirePro V7800, discussed in our previous paper, had multiple VLIW units arranged in a SIMD design.

The Arithmetic Logic Unit (ALU) performs basic arithmetic operations. Often ALU without further descriptors refers to integer arithmetic operations specifically. A Floating point unit (FPU) is an ALU that performs arithmetic operations with floating point formats. FPUs typically work with the single and double precision formats from the \textit{IEEE Standard for Floating-Point Arithmetic 754} \cite{IEEE754}. The processing elements of a GPU are essentially FPUs. When GPGPU first became available they were not fully compliant with the floating point specification, but recent models claim compliance.

GPUs operate as a collection of independent SIMD-like units, so they operate well with the OpenCL model. CPUs typically have a SIMD ability, though generally on a much smaller scale than on a GPU. A primary difference of interest is the CPU's focus on data/intra-thread locality while the GPU focusses on inter-thread locality. That is, on a CPU, data for one thread is logically co-located into a contiguous block of memory, but for a GPU the memory is ordered to suit a single memory access by each thread concurrently. With a symmetric (filled) block tridiagonal matrix, for example, it is trivial to experiment with both of these methods to see which has the best performance; one simply has to switch the blocks and change minor piece of calculation logic.

C, specifically the C99 standard, was chosen over C++ and Fortran. The chief reason for choosing C99 was that OpenCL C is a C99 variant, and the OpenCL specification is primarily defined as a C interface. As a language, C is conceptually simple with a small amount of syntax to learn, yet very powerful. The heavy numerical processing in this case was performed by OpenCL C code; the advantages within Fortran for numerical processing are thus not relevant.

In the following sections we lay out the necessary details for the ab-initio calculation of hydrogenic systems under intense linearly polarized laser fields. In section \ref{sec:TDSE} a description of the underlying theory is given, followed by section \ref{sec:EM} which discusses the form of the electromagnetic fields. The mathematics for calculating the observables and gauge dependent quantities are discussed in section \ref{sec:Obs}. The remainder of the paper focusses on the specifics of installing (section \ref{sec:Install}), analysing (section \ref{sec:Code}), using (section \ref{sec:User}), and modifying (section \ref{sec:Mod}) the code. An overview of the settings is provided in \ref{sec:Conf}, and the algorithms for the Lanczos (\ref{sec:Lanczos}) and Runge-Kutta approaches (\ref{sec:RungeKutta}) are also given to give further context.

\section{Finite-difference formulation of the time-dependent Schr\"{o}dinger Equation} \label{sec:TDSE}

The field-free Hamiltonian of the atomic system $\hat{h}(\v{r})$ reads,
\begin{equation}
\hat{h}(\v{r}) = -\frac{1}{2}\nabla^{2} + V(r),
\label{eq:h_molecular_ion}
\end{equation}
where $\v{r}$ is the position vector and the spherically symmetric potential $V(r)$ is assumed to satisfy $rV(r\rightarrow \infty) \longrightarrow const $. For example, for pure hydrogenic systems, the potential is given as $ V(r) = -Z_{nuc}/r$, where $Z_{nuc}$ is the atomic number.

The Hamiltonian of this system in a linearly-polarised radiation field $\v{E}(t)=\hat{e}\mathcal{E}(t)$, where $\hat{e}$ is the polarisation vector, can be expressed in different forms depending on the chosen gauge used to represent the atom-field interaction operator. In the present implementation we have adopted the length (L) and the velocity (V) forms in the dipole approximation, expressed as:
\begin{equation}
\hat{d}^{(G)}(\v{r},t) =
\left\{ \begin{array}{cl}
 \frac{1}{c}{\v{A}}(t)\cdot {\v{p}} &\qquad  G = V \\
 \v{E}(t) \cdot \v{r}, &\qquad  G = L
\end{array} \right.
\end{equation}
where $\v{p} = -i \nabla $ represents the momentum operator and $\v{A}(t)=-c\int_{-\infty}^t dt^\prime \v{E}(t^\prime)$ is the vector potential of the field.

In this case, the Hamiltonian becomes gauge dependent, since $\hat{H}^{(G)}(\v{r},t) = \hat{h}+\hat{d}^{(G)}(\v{r},t)$, with the time evolution of the system's wavefunction which is given by,
\begin{equation}
i\frac{\partial}{\partial t}\psi_G(\v{r},t)= \left[ \hat{h} + \hat{d}^{(G)}(\v{r},t) \right] \psi_G(\v{r},t).
\label{eq:tdse}
\end{equation}
Based on the above, it is concluded that the time-dependent wavefunction also becomes gauge dependent and it is known that the two forms of the wavefunction are related through the gauge transformation:
\begin{equation}
\psi_L(t) = e^{ -i \v{A}(\v{r},t)\cdot \v{r}}\psi_V(t)
\end{equation}

In our present numerical implementation we choose an orthogonal coordinate system with the $z-$axis along the polarisation vector of the field; $\hat{z}=\hat{e}$. Moreover, although in principle a partial wave expansion of the wavefunction should include an infinite summation of the quantised angular momentum terms, in an actual calculation we truncate the expansion of the spherical harmonics at some maximum $l = l_{max}$. The maximum angular momentum $l_{max}$ should be chosen such that the maintenance of unitarity during time propagation is ensured and the expectation values of the observables under study have sufficiently converged as $l_{max}$ increased. The time-dependent wavefunction $\psi_G(\v{r},t)$ is expanded on a basis of spherical harmonics $Y_{lm_l}(\theta, \phi)$:
\begin{equation}
  \psi_G(\textbf{r},t)=
  \sum_{l, m_l}\frac{1}{r}f^{(G)}_{l,m_l}(r,t)Y_{lm_l}(\theta,\phi), \qquad G=L,V
\label{eq:wf_lm}
\end{equation}
where the partial waves $f^{(G)}_{l,m_l}(r,t)=r\langle lm_l|\psi_G\rangle$ are the time dependent quantities to be evolved in time.  We proceed by the substitution of equation \ref{eq:wf_lm} into equation \ref{eq:tdse}, followed by a projection onto the spherical harmonic conjugates $Y^{\star}_{lm_l}(\theta,\phi)$ and integration over the solid angle $d\Omega = \sin \theta d\theta d\phi$. After employing the orthnormalisation properties of the spherical harmonics, we arrive at the following propagation scheme for the partial waves \cite{KulanderEtAl1992} \cite{NikolopoulosEtAl2008}:
\begin{equation}
 i\frac{\partial}{\partial t} f^{(G)}_{l,m_l}(r,t) = \hat{h}_{lm_l}(r) f_{l,m_l}(r,t) + \sum_{l^\prime m_{l^\prime}} \hat{d}^{(G)}_{lm_l,l^\prime m_{l^\prime}}(r,t) f^{(G)}_{l^\prime , m_{l^\prime}}(r,t),
\label{eq:tdse_radial}
\end{equation}
where the diagonal $\hat{h}_{lm_l}$ and non-diagonal $d^{(G)}_{lm_l;l^\prime m_{l^\prime}}$ radial operators are defined
by
\begin{eqnarray}
 \hat{h}_{lm_l}(r) &=&  \langle lm_l|\hat{h}(\v{r})|l^{\prime}m_{l^{\prime}}\rangle =
\left[
-\frac{1}{2}\frac{\partial^{2}}{\partial r^{2}} + \frac{l(l+1)}{2r^{2}} + V(r)
\right]
\delta_{ll^{\prime}}\delta_{mm^{\prime}},
\label{eq:tdse-radial-field}\\
\hat{d}^{(G)}_{lm_l;l^{\prime}m_{l^{\prime}}}(r,t) & = & \langle lm_l|\hat{d}(\v{r},t)|l^{\prime}m_l^{\prime}\rangle
= \hat{t}_{ll^\prime}^{(G) }(r,t) k_{l,l^\prime}(m_l)
\\
k_{l,l^\prime}(m_l) &=& \langle l m_l | \cos \theta | l^\prime m_{l^\prime}  \rangle = \, \delta_{ll^\prime \pm 1}\delta_{m_l m_l^\prime}\sqrt{ \frac{l^2_> - m_l^2}{4l^2_> -1}}
\label{eq:tdse-potential}
\end{eqnarray}

with $l_> = \max(l,l^\prime)$ and $\hat{t}_{ll^\prime}^{(G)}$ the radial coupling operator of each gauge:
\begin{equation}
\hat{t}^{(G)}_{ll^\prime}(r,t) =
\left\{
\begin{array}{cl}
   -\frac{i}{c}A(t)\left[\frac{\partial}{\partial r}+ (l-l^\prime)\frac{l_{>}}{r}\right] , &\quad G = V\\
   r\mathcal{E}(t), &\quad G=L
\end{array}\right.
\end{equation}
As seen from the above expressions, the electromagnetic interaction operator couples only states with equal magnetic quantum number, $m_l= m_{l^\prime}$ and angular momentum numbers that
differ by one unit, $l^\prime = l\pm1$. Therefore the TDSE in the two gauges, length and velocity are finally written as:
\begin{eqnarray}
i\pd{}{t} f^{(L)}_{l,m_l}(r, t) &=& \hat{h}_{lm_l}(r)f^{(L)}_{l,m_l}(r, t) + r\mathcal{E}(t) \left[ \kappa_{l,m_l} f^{(L)}_{l-1,m_l}(r, t) + \kappa_{l+1,m_l} f^{(L)}_{l+1,m_l}(r, t) \right] \\
i\pd{}{t} f^{(V)}_{l,m_l}(r, t) &=& \hat{h}_{lm_l}(r)f^{(V)}_{l,m_l}(r, t) - \\ \nonumber
 && i \frac{A(t)}{c} \left[ (\pd{}{r} + \frac{l}{r}) \kappa_{l,m_l} f^{(V)}_{l-1,m_l}(r, t) + (\pd{}{r} - \frac{l+1}{r}) \kappa_{l+1,m_l} f^{(V)}_{l+1,m_l}(r, t) \right ],
\end{eqnarray}
with $\kappa_{l,m_l}=k_{l,l-1}(m_l)$ and $\kappa_{l+1,m_l}=k_{l,l+1}(m_l)$.

Up to this point the derivation of the time-dependent equations is general and no reference to any particular integration scheme has been made. One choice is the expansion of the partial waves $f^{(G)}(r,t)$ on some known basis, which will transform the above partial-differential equation into a coupled system of ordinary differential equations for the expansion coefficients \cite{BachauEtAlRepProgInPhys2001}. This approach is known as the basis representation method of the time dependent wavefunction. In the present case we have chosen the grid representation of the wavefunction where the quantities that are treated numerically are the values of the partial wave functions $f^{(G)}(r,t)$ on a specific radial grid. Considering this grid based approach, the differential operators, which appear in the TDSE, are expressed through a finite difference scheme. In general there are two different ways to conceptualise a grid based problem: in terms of the individual elements communicating between neighbours, or in terms of a linear algebra problem where the derivative is represented by a matrix vector calculation and where the primary calculation is that of an exponential operator acting on a vector. In the present formulation the latter approach is used for the time evolution of the system. An evenly spaced grid is chosen from the origin up to a distance R, which represents the spatial region where the system is located. Denoting the grid as $r_j =(j-1)h, j=1,N$, where $h=R/(N-1)$ is the constant grid spacing, and dropping the indexes $l,m_l$ temporarily, then the partial wave functions and their spatial derivatives are expressed as
\begin{eqnarray}
f_{j}(t) &=& f^{(G)}_{l,m_l}(r,t)   \\
\frac{d}{dr}f_{j}(t) &=& \frac{1}{12h^2} [-f_{j+2}(t) + 8f_{j+1}(t) - 8f_{j-1}(j-1) + f_{j-2}(t) ] + O(h^4) \\
\frac{d^2}{dr^2}f_{j}(t) &=& \frac{1}{12h^2} [-f_{j+2}(t) + 16f_{j+1}(t) - 30 f_{j}(t) + 16f_{j-1}(t) -f_{j-2}(t)]+ O(h^4)
\end{eqnarray}
where the first and the second derivative have approximated with fourth order difference formulae.

\begin{figure}[!t]
\centering
\includegraphics[width=180px, height=180px]{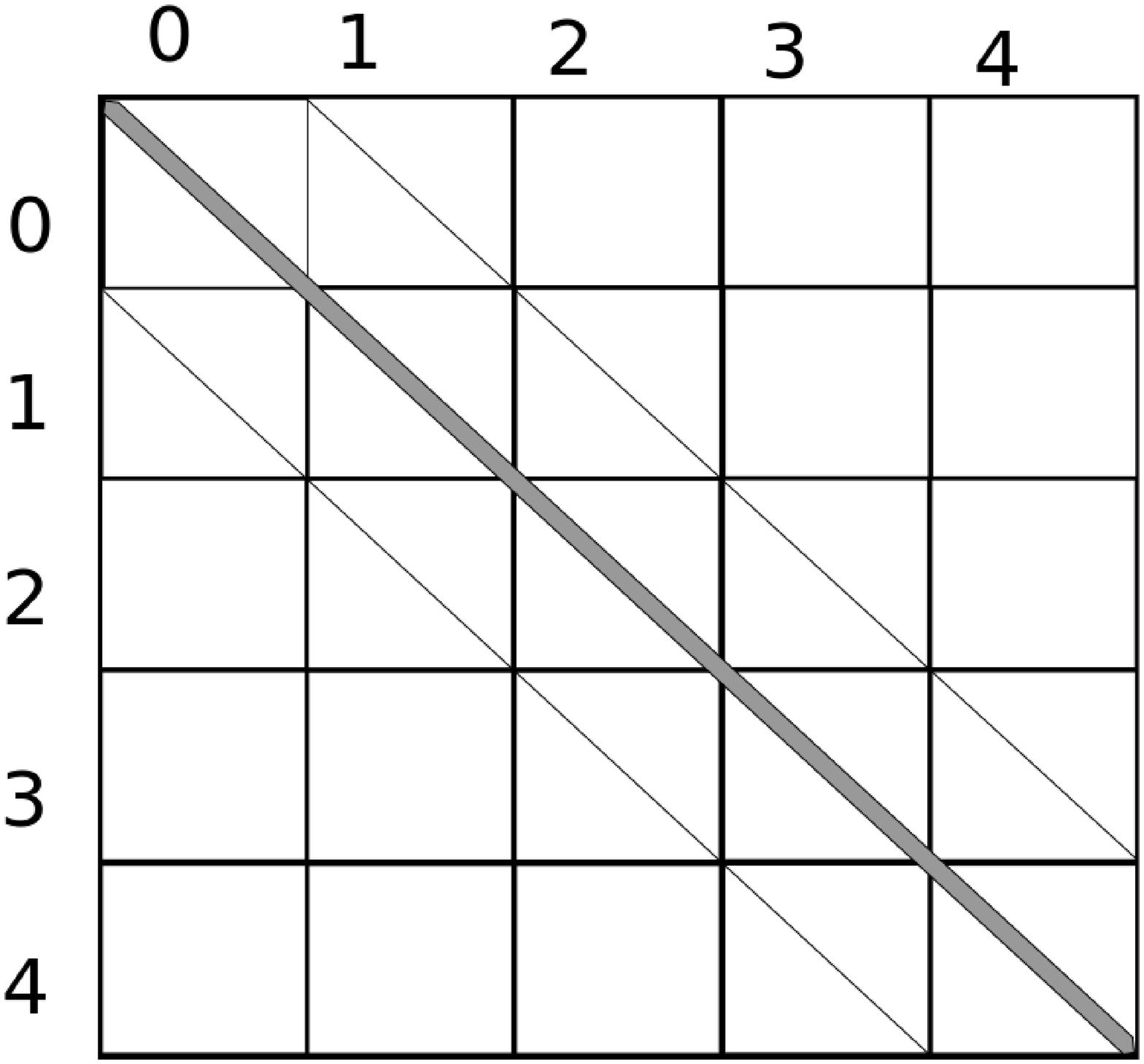}
\hfill
\includegraphics[width=180px, height=180px]{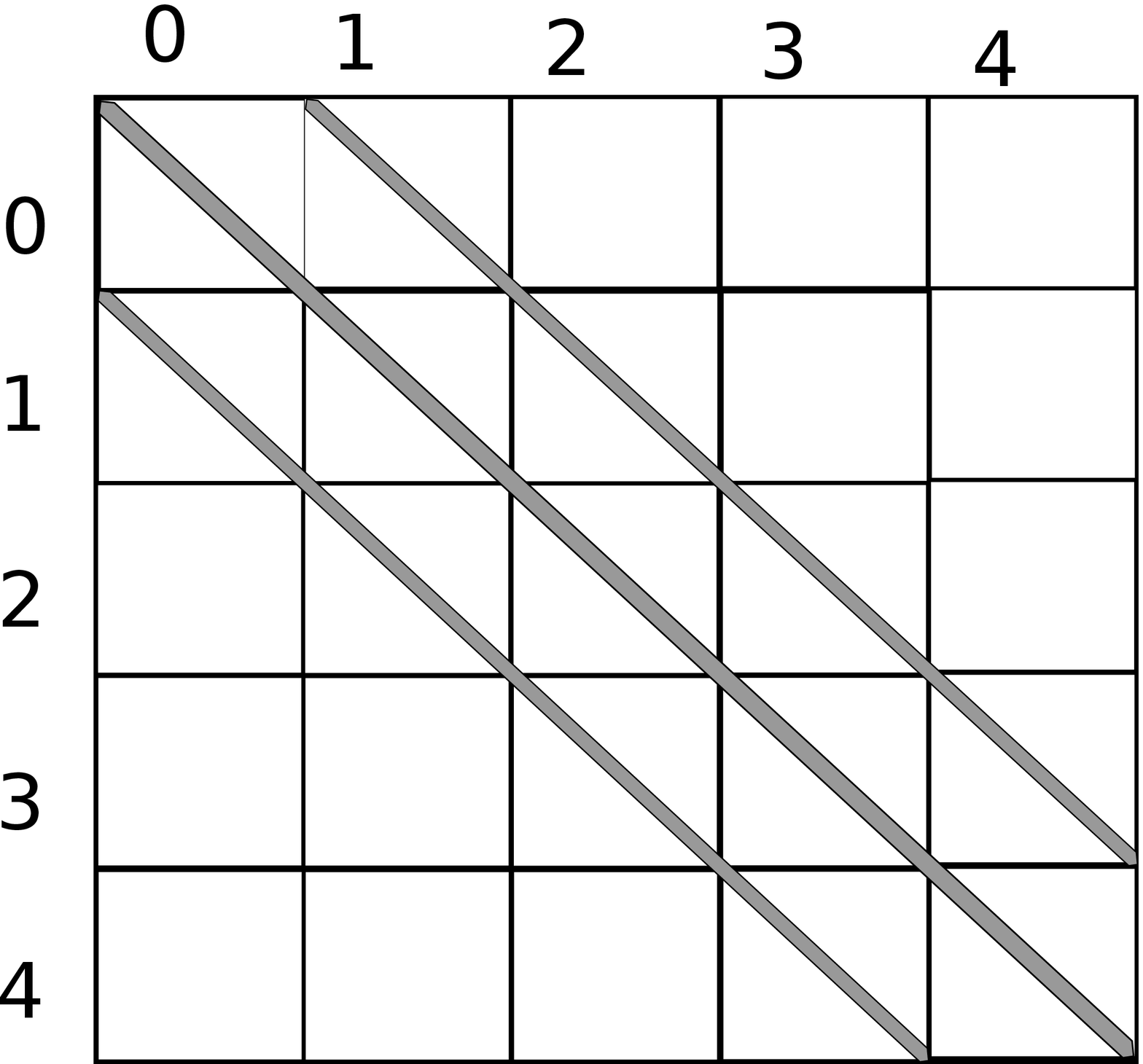}
\caption{The banded structure of the finite difference Hamiltonian in the length gauge (left figure) and the velocity gauge (right figure) for $l_{max}=4$  is shown. For the case of the velocity gauge, the sub diagonal, super diagonal and diagonal blocks are all multi-diagonal to account for the particular finite difference scheme used for the first and second derivatives.
}
\label{fig:FDLength}
\end{figure}

Formally, the system of equations is written in a compact way by using matrix notation. Since the magnetic quantum number index $m_l$ is a constant of the motion, we may define a vector such that $\v{F}_{m_l}^{(G)}(t) = \{[f^{(G)}_{0,m_l}(r_1,t),..., \\ f^{(G)}_{0,m_l}(r_j, t),..., f^{(G)}_{0,m_l}(r_N,t)],....,[f^{(G)}_{l_{max},m_l}(r_1,t),....,f^{(G)}_{l_{max},m_l}(r_j,t),...,f^{(G)}_{l_{max},m_l}(r_N,t)]\}$ and the matrix,
\begin{equation}
\v{H}_{m_l}^{(G)}(t) =
\left[
\begin{array}{ccccc}
\v{h}_{0}       &  \v{d}^{(G)}_{01}(t)    &  0                  &  ...       &          0            \\
\v{d}^{(G)}_{10}(t)      &  \v{h}_{1}     & \v{d}^{(G)}_{12}(t)   &  ...       &          0            \\
    0                &  \v{d}^{(G)}_{21}(t)  &  \v{h}_{2}     &  ...       &         ...           \\
   ...               &     ...             &  ...                &  ...       &         ...           \\
   ...               &     ...             &  ...                & \v{h}_{l_m-1} &  \v{d}^{(G)}_{l_m-1,l_m}(t)  \\
    0                &     0               &  ... & \v{d}^{(G)}_{l_m,l_m-1}(t)       &      \v{h}_{l_m} \\
\end{array}
\right],
\label{eq:h_matrix}
\end{equation}
with $\v{h}_l$ and $\v{d}^{(G)}_{ll^\prime}$ being the $N \times N$ matrix representations of the operators $\hat{h}_l(r)$ and  $\hat{d}^{(G)}(r,t)$ on the chosen finite-difference scheme respectively. In this case the time-dependent coupled equations Eqns. \ref{eq:tdse_radial} are expressed in their final matrix-representation form as:
\begin{equation}
  \dot\v{F}_{m_l}^{(G)}( t) =-i\v{H}_{m_l}^{(G)}(t) \v{F}_{m_l}^{(G)}( t).
\label{eq:tdse_matrix}
\end{equation}
The structure of the propagation matrix $\v{H}_{m_l}^{(G)}$ for the length and the velocity gauge is shown in Figs. \ref{fig:FDLength}.

Note that this block structure is similar to the block structure obtained with a basis representation of the wavefunction as described in \cite{OBroinNikolopoulos2012} (see Fig. 2). The difference with the present finite-difference formulation is that now the diagonal blocks will not be diagonal themselves but be multi-diagonal. The off-diagonal blocks themselves are now diagonal (length gauge) or multi-diagonal (velocity gauge), while in the basis representation case \cite{OBroinNikolopoulos2012} they were dense.

\subsection{Absorbing potential}
Up to this point, the formulation has been presented by artificially placing the atomic system inside a sphere of radius $R$, by forcing the wave function to be zero after the fixed boundary. That is, the terms for the finite difference are assumed to be zero after the fixed boundary, much like in the case of an infinitely high box potential. This might be a source of artificial complications, such as the unphysical reflection of parts of the time-dependent wave function at the boundaries. To overcome this problem, an absorbing boundary has been added throughout the inner region of the box. This is implemented through the addition of an imaginary potential into the Hamiltonian, through the substitution $ V(\v{r})\rightarrow V(\v{r}) + W(r)$ , where $W(r)$ is \cite{NikolopoulosEtAl2007}

\begin{equation}
W(r)=\frac{i}{dt}\ln \left\{1 - \cos^{p}\left[\frac{\pi}{2}\left(1-\frac{r}{r_{b}}\right)\right] \right\},
\end{equation}
where $p$ controls the \textit{smoothness} of the imaginary potential. Absorption is, essentially, absent at the central region of the system, $\lim_{r \to +0} W(r) = 0$, while it approaches complete absorbtion at the box boundaries, $\lim_{r \to r_b} W(r) = -i\infty$.

The imaginary potential not only solves the technical problem of the artificial scattering but it can also be used as a method to describe the ionisation process. This imaginary potential smoothly removes the flux which approaches the boundaries, affecting only the continuum part of the wave function. Therefore, the loss of norm of the wave function could be interpreted as ionisation of the system, although in practice this should be used in conjunction with the spatial integration method of calculating the yield, which was discussed earlier, so that the yield calculation can be done practically without waiting for all continuum parts of the wavefunction to reach the box boundaries.
\section{Electric field and vector potential} \label{sec:EM}

In the present implementation the electric field $\mathcal{E}(t)$ and the vector potential $A(t)$ of the pulse are represented as
\begin{eqnarray}
  \mathcal{E}(t) &=& \mathcal{E}_0 f(t) \cos(\omega_0 t + \phi_{cep}(t)) \label{eq:EField},\\
  A(t) &=& - c \int_{t_0}^{t} d t^\prime \mathcal{E}(t^\prime),
\end{eqnarray}
where $\omega_0$ is the photon energy, $\mathcal{E}_0$ is the maximum of the envelope, $f(t)$ is the pulse envelope and  $\phi_{cep}$ is the carrier-envelope phase, which is the phase difference between the carrier wave and the envelope. Since we have adopted the dipole approximation, it is assumed that both the electric field and the vector potential are approximately constant over the extent of the box at some instant in time; that is, the spatial variation of the fields are ignored \cite{BransdenJoachain2003}.

At present two forms of the envelope are implemented the Hann function envelope and the Gaussian envelope.

The Hann function envelope (sine squared pulse) adopted is given by
\[
f(t) = sin^2(\frac{\omega_0}{2n_c}),
\]
where $n_c$ is the number of cycles of the carrier wave.

In particular, for the Gaussian pulse we have also implemented the case where the electric field phase is not necessarily constant, i.e the pulse can be chirped:

\begin{eqnarray}
f(t) &=& \frac{E_0 z_0}{z_d^{\frac{1}{4}}} e^{ - \frac{z_0^2 t^2}{2z_d}} \cos \left (\omega_0 t + \phi_{cep}+\frac{d}{2z_d} t^2 - \frac{\arctan(d/z_0^2)}{2} \right ),\\
z_0 & = & \frac{\tau_d}{2\sqrt{2 \ln 2}},
\qquad
 z_d^2  = z_0^4 + d^2.
\end{eqnarray}
For a pulse without chirp $d=0$, $\tau_d$ represents the FWHM of the pulse intensity, and the above equatio simplifies to
\begin{equation}
f(t) = E_0 e^{\left ( -2\sqrt{2 \ln(2)} \frac{t^2}{\tau_d^2} \right )} \cos(\omega_0 t)
\end{equation}

Having chosen the form of the electric field $E(t)$ we numerically calculate the potential field $A(t)$, for each time step, as:
\[
A(t_{n+1}) = A(t_{n}) - c \int_{t_{n}}^{t_{n} + h} d\tau E(\tau), \label{eq:VectorStep}
\]
where the integral is performed using a standard 5-point Gaussian quadrature. The speed of light factor, c, does not need to be explicitly included in the actual code since there is a $\frac{1}{c}$ factor in the transition operator.

For example, for a potential field $A(t)$ with a Hann function shape, the electric field has the form:
\[
    E(t) = E_0 \sin (\Omega t) \left( \omega_0 \sin (\Omega t) \cos (\omega_0 t) + 2 \Omega \cos (\Omega t) \sin(\omega_0 t) \right ) \label{eq:VecSinSinCosE}
\]
where $\Omega_0 = \frac{\omega_0}{2 n_c}$. This can be viewed as a combination of two electric fields which are individually of the form given in Eq. \ref{eq:EField}. The vector potential in the velocity gauge is thus given by
\[
  A(t) = -\frac{E_0}{c} \sin^2\Omega t \sin\omega_0 t
\]

Although the Gaussian envelope represents a very realistic description of the actual experimental fields, the Hann form of the envelope is used very frequently as it has some numerical advantages against the Gaussian envelope. First, in contrast to the Gaussian envelope, which is non-zero for all times, the Hann envelope is strictly zero at precisely known end points $t_i = 0$ and $t_f=\pi/(n \omega_0)$. In addition, the  Fourier transform of the Hann envelope results in a sharper cut-off in the spectral distribution for frequencies other than the carrier frequency. This is a very important property for few-cycles pulses where bandwidth effects can have appreciable effects on the final results. This is particularly true when the photon energy is near resonant with transition energies of the system. Finally, in comparison with a Gaussian pulse with the same total electromagnetic energy offered in the system, there is an appreciable amount of computing time that is saved because the extent of the pulse is more limited.

\section{Calculation of observables} \label{sec:Obs}

\subsection{Ground state calculation}
The \textit{diffusion} equation may be used to locate the eigenstates and associated eigenenergies for a stationary system. In this case, we assume that no field is present and the corresponding Time Independent Schr\"odinger Equation (TISE) is further modified by performing the replacement $-i \rightarrow 1$  \cite{FlocardEtAl1978}. The resulting partial differential equation is the well known \textit{diffusion equation}. The method is particularly effective for finding the lowest energy eigenstate of an uncoupled state subspace as it does not require information about any other eigenstate. The replacement $-i \rightarrow 1$ breaks the ability of the state evolution operator to maintain unitarity. The partial waves $f_{lm_l}(r,t)$ are re-normalized after each time step to ensure that the wavefunction does not grow overly large and compromise numerical accuracy. The magnitude of the break from unitarity can be used to calculate the expectation value of the Hamiltonian operator, through the following expression \cite{NikolopoulosEtAl2007}:

\begin{eqnarray}
\avg{E(t)}dt &=& - ln \braket{\psi(t+dt)}{\psi(t+dt)} 
\label{eq:e_g}
\end{eqnarray}


It is important to note that the method does not find the true physical ground state, but instead it finds the ground state corresponding to the discretised numerical system. If we start with the analytical ground state, the system will quickly approach the ground state of the numerical approximation of the system. If the time step used is too large, numerical accuracy will be compromised and the algorithm may not converge on the ground state. If the time step is too small, more computational time than is necessary may be used in reaching the ground state. Using the time steps of the standard time dependent calculation provides reasonable values.

It is also important to bear in mind that the symmetry properties of the initial state also determine which energy eigenstate the system settles into. For example, if only states of a particular angular momentum $l$ are initially populated then the system can only settle into the lowest energy eigenstate with the same angular momentum $l$, provided that there are no non-field dependent couplings among different partial waves (as for example in the molecular case). This is because there is no field to couple states which differ in angular momentum.

\subsection{Populations}
\begin{figure}[!t]
\centering
\includegraphics[width=200px]{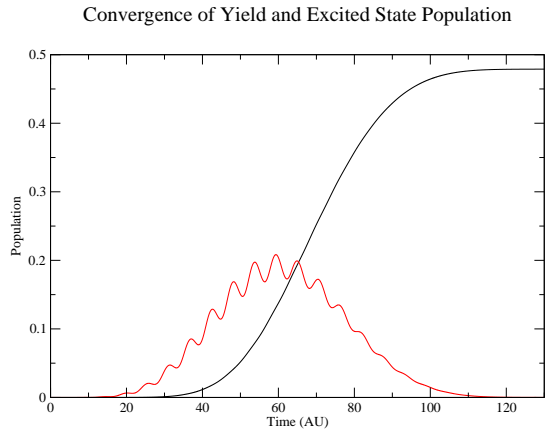}
\caption{ The sine-squared vector potential returns to zero after approximately 140 au. The dashed red line is the calculated yield, while the solid black line is the calculated excited state populations. Only the final population values are physically meaningful; the ionised population requires time post-propagation to be radially separated from the excited state population. The yield is calculated with an inner boundary that ignores the first 50 finite difference grid points ($r_{Ion}$). The excited state population is calculated by subtracting the ground state population and the yield from unity. }
\label{fig:Yield}
\end{figure}
During and after the interaction of the atomic system with the radiation field, the vector of the radial wavefunctions $f_{lm_l}(r,t)$ 
are used within the finite difference approach and so can also be used for the calculation of other quantities. Having the data for the evolution of the radial wavefunctions to hand, the most straightforward quantities to calculate are the ground state population, the excited state population and the total ionisation yield.

The ground state population $p_g(t)$ is calculated by the projection of the initial state $\psi(\v{r},0)$ onto the wavefunction at time $t$, $\psi(\v{r}, t)$,
\begin{equation}
p_g(t) = \braket{\psi(\v{r}, 0)}{\psi(\v{r}, t)} = \int_{0}^{R} dr f^\star_{0,0}(r, 0) f_{0,0}(r, t).
\label{eq:p_g}
\end{equation}

The population of the bound states can be calculated by direct spatial integration of the probability values at all grid points inside a chosen radius, say $r_b$:
\begin{equation}
p_b(t) = \langle \psi(t)|\psi(t)\rangle_{r<r_b} =
\sum_{l m_l}^{l_{max}} \int_{0}^{r_b} dr |f_{l,m_l} (r, t)|^2.
\label{eq:p_b}
\end{equation}

The population of the excited states, $p_e(t)$, can be deduced by knowing the ground state population Eq. \ref{eq:p_g} and the bound state population Eq. \ref{eq:p_b}  as $p_e(t)=p_b(t)-p_g(t) \label{eq:p_e}$.

The two methods of calculating the yield, below, require post-pulse propagation. This means that the ionised population will be counted as the excited state population during the run of the pulse. That is, the excited state population is estimated by the quantity within a subset of the box which is not counted towards ionisation, and which is not the ground state. As such, one estimates the excited states population as $p_e(t)=p_b(t)-p_g(t)$. Note that it is not until the post propagation, when the ionised population moves away from the central potential and the yield value asymptotically approaches a value as in Fig. \ref{fig:Yield}, that the excited state population becomes meaningful.

\subsection{Ionisation yield and angular distribution}
In the present case we use two different methods to calculate the ionisation yield; (a) direct spatial integration of the probability values at all grid points outside a chosen radius, say $r_b$, and (b) use of the probability current directed radially through a sphere of radius $r_b$, as shown later.

\paragraph{Spatial integration of the wavefunction}
Assuming spatial integration for radius $r>r_b$ we obtain
\begin{equation}
p_i(r_b,t) = \langle \psi(t)|\psi(t)\rangle_{r>r_b} = \sum_{l m_l}^{l_{max}} \int_{r_b}^{R} dr |f_{l,m_l} (r, t)|^2 =
1 -  \sum_{l m_l}^{l_{max}} \int_{0}^{r_b} dr |f_{l,m_l} (r, t)|^2.
\label{eq:yield_1}
\end{equation}

To ensure that a reasonable part of the contributions from continuum energy-eigenstates are counted in the ionisation yield post-calculation, the wave equation is propagated forward in time. This allows the continuum contributions to move away from the central boundary so that the yield can be calculated by counting the probabilities after a certain cut-off radius $r_{b}$. This calculation is only valid when the complex potential term is not used to siphon off the probability current heading towards the boundaries. If the complex absorbing potential term is present, the break from the norm should also be added to the yield. When no absorbing potential is present then
\begin{equation}
p_i + p_b = 1, \label{eq:Norm}
\end{equation}
assuming perfect numerical accuracy.

\paragraph{Probability current}
\begin{figure}[!t]
\centering
\includegraphics[width=400px]{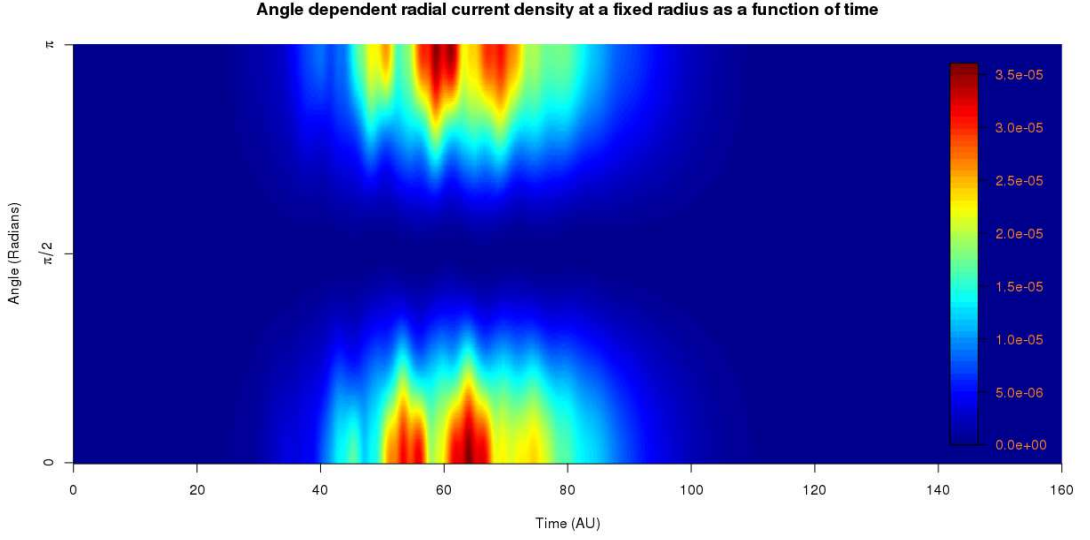}
\caption{The strength of the time-dependent radial probability current $j_r(r_b,\Omega, t)$ at a fixed radius $r_b = 9.98$ au and for $\theta$ from  $0$ to $\pi$, over the duration of the pulse described in Fig. \ref{fig:VecPot}. Since $m=0$, the angle, $\phi$, plays no role. } \label{fig:RadCurr}
\end{figure}
The radial probability current of the time-dependent wave function is obtained from the probability current $\v{j}(\v{r}, t)$:
\begin{eqnarray}
& & j_{r}( \v{r},t) = \hat{\v{r}}  \cdot \v{j}(\v{r}, t) = \hat{\v{r}} \cdot {\rm Re} \left\{ \psi^\star(\v{r},t) \left[ \v{p} + \frac{1}{c}\v{A}(t) \right] \psi(\v{r},t)\right\} \\ & &= - {\rm Im} \left\{ \frac{1}{r} \sum_{lm_l} \sum_{l^\prime m_l^\prime} f^\star_{l,m_l}(r,t) \pd{}{r} \left [ \frac{1}{r} f_{l^\prime, m_l^\prime}(r,t) \right] Y^\star_{lm_l}(\Omega) Y_{l^\prime m_l^\prime}(\Omega) \right\} \\ & & + \frac{1}{r^2} \frac{A(t)}{c} \cos(\theta) \left | \sum_{l m_l} f_{l,m_l}(r,t) Y_{lm_l} (\Omega) \right |^2 \\ \nonumber \label{eq:flux}
\end{eqnarray}
with the momentum operator $\v{p} = -i\nabla$ and the unit radial vector $\hat{r}= \v{r}/r$ where $\v{r}=(r,\Omega)$. By use of the continuity equation ($\nabla \cdot \v{j}(\v{r},t) + \partial \rho(\v{r},t)/\partial t = 0$), it follows that $j_{r}(r_b,\Omega,t)r_b^{2} d\Omega$ provides the number of electrons moving outwards, per unit time, within a solid angle $d\Omega$ through a spherical surface placed at some distance $r_b<R$. For the differential probability we have:
\begin{equation}
 \frac{dP}{d\Omega}(r_b,\Omega,t)
= \int_{0}^{t}dt^\prime r_b^{2} j_{r}(r_b,\Omega,t^\prime).
\label{eq:pad}
\end{equation}

The time $t$ is chosen to be large enough such that all radial outgoing flux has passed the point of observation $r_b$. The distance $r_{b}$ is chosen neither too close to the central region nor to the box boundaries. This is to ensure two things respectively, that no flux corresponding to the system moving into an excited state is included and so the radial distance at which the absorbing potential $W(t)$ has a significant, descernible impact if it is enabled. Where the absorbing potential is non-zero, the continuity equation is modified as $\nabla \cdot \v{j}(\v{r},t) + \partial \rho(\v{r},t)/\partial t = 2W(t)\rho(\v{r},t)$. We make the choice of $r_b$ in order to avoid any complications introduced by the modification of the continuity equation.

For a fixed radius $r_b$, the quantity $j_r(r_b, t)$, shown in Fig. \ref{fig:RadCurr}, can be used to calculate the ionisation current. It is self evident that the square of the time integral of the probability current flowing through the sphere of radius $r_b$ is exactly the population outside the sphere, considering that the initial population outside the sphere is effectively zero. For this method, adding the break from the norm is not required. Therefore, by integrating  Eq. \ref{eq:pad} over the angular variables we obtain the ionisation yield $p_i(t)$ at time $t$ as,
\begin{equation}
p_i(r_b,t)=\int_{0}^{t}dt^{\prime}\,
\sum_{l,m_l}  \left\{ {\rm Im} \left[ f^\star_{l,m_l}(r_b,t^\prime) \frac{\partial }{\partial r} f_{l,m_l}(r_{b},t') \right] + \frac{2}{c}A(t) \kappa_{l+1,m_l}f^\star_{l,m_l}(r_b,t^\prime) f_{l+1,m_l}(r_b,t^\prime) \right\}. \label{eq:yield}
\end{equation}
The above quantity, over the lifetime of the simulation, provides a separate measure of the ionisation yield.

\subsection{High harmonic generation}
\begin{figure}[!t]
\centering
\includegraphics[width=300px]{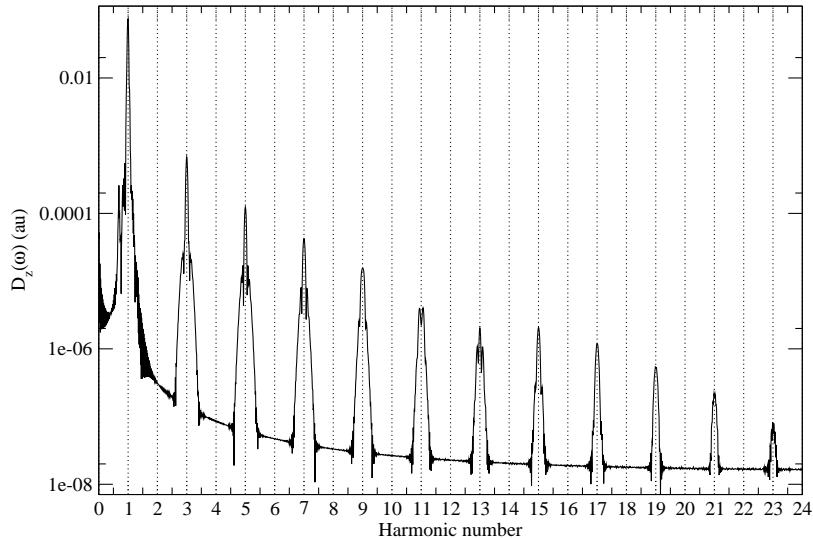}
\caption{ The Fourier transform of the dipole moment for a hydrogen system under the action of a pulse of intensity $1\e{15}$ W/cm$^{2}$, central photon energy $15$ eV, 40 cycles and with a sine-squared shape for the vector potential. Multiples of the photon frequency, on the x axis, are plotted again the dipole moment strength in atomic units on the y axis.}
\label{fig:HHG}
\end{figure}
In the particular case where $m_l=0$, we also calculate the expectation value of the dipole moment, dipole velocity, and dipole acceleration, using the following formulae \cite{BandraukEtAl2009}:
\begin{eqnarray}
 \avg{z(t)} &=& 2 \sum_{l} \kappa_{l+1,0} {\rm Re}  \int_{0}^{R} dr r f^{\star}_{l,0} (r, t) f_{l+1,0} (r, t)  \label{eq:Dipole}\\
 \avg{\dot{z}(t)} &=& \sum_{l} \kappa_{l+1,0} {\rm Im} \\ \nonumber & &
 \left[ \int_{0}^{R} dr  r f^{\star}_{l+1,0} (r, t) \pd{}{r} \left( \frac{1}{r} f_{l,0}(r,t) \right) - l \int_{0}^{\infty} dr f^{\star}_{l+1,0} (r, t) \frac{1}{r} f_{l,0} (r, t) \right] \\ \label{eq:DipoleVelocity}
 \avg{\ddot{z}(t)} &=& - E(t) - 2 \sum_{l} \kappa_{l+1, 0} {\rm Re}  \int_{0}^{R} dr \frac{1}{r^2} f^\star_{l,0} (r, t) f_{l+1,0} (r, t) \label{eq:DipoleAccel}
\end{eqnarray}

Since the wavefunction is set so as to be finite in extent, the integration can be terminated at the end of the box $R$ instead of at infinity. Figure \ref{fig:HHG} is an example plot that clearly shows odd harmonics spaced by twice the photon frequency.

\section{Description of the code}\label{sec:Code}

OpenCL is a specification that attempts to standardize the use of co-processors. The current version of the standard is 1.2  (Rev 19), and was released by the Khronos OpenCL group, the multi-vendor committee responsible for the specification. It was at the suggestion of Apple\textsuperscript{\textregistered} that the initial specification came about.

\prog is provided as a standard program, and not a library.

\paragraph{Code reusability and extensibility} The overarching concept of the code's framework design is \textit{reuseability} of all functions. That is, the functions that are lowest on the function tree should also be the most generic in their operation. Functions should have narrowly defined, but generically implemented, functionality. Tasks such as the opening of files, or the allocation of memory, are generally passed to functions higher up the tree where reasonable.

\textit{Extensibility} is also important because not all possible use cases can be anticipated, and the alteration and generation of code at a high level up the function tree is desired, That is, lower level functionality should be able to be called upon in unforeseen circumstances, but within the scope of the function, and still be expected to work.

For example, \prog was initially written for the basis method \cite{OBroinNikolopoulos2012}, but was extended to duplicate calculations for the TDDM system described in \cite{NikolopoulosEtAl2011}, and also to work for the present finite difference implementation. The intention of the framework is for it to be suitable for use cases which rely on a single compilation of the OpenCL code. It has not been tested for use cases involving repeated initialisation and deallocation of the framework.

With initialisation, performance considerations are generally not a primary concern, as almost all of the total running time is dictated by the specific computation. Only in the smallest computations does initialisation bear any reasonable performance penalty. As such, in very short but repeated computations, tailoring specific code to handle the repeated computation is more desirable than scripted re-runs of the program. Scripted re-reruns will involve repeated initialisation and repeated compilation. The extensibility of the code plays an important factor here.

\begin{figure}[!t]
\centering
\includegraphics[width=400px]{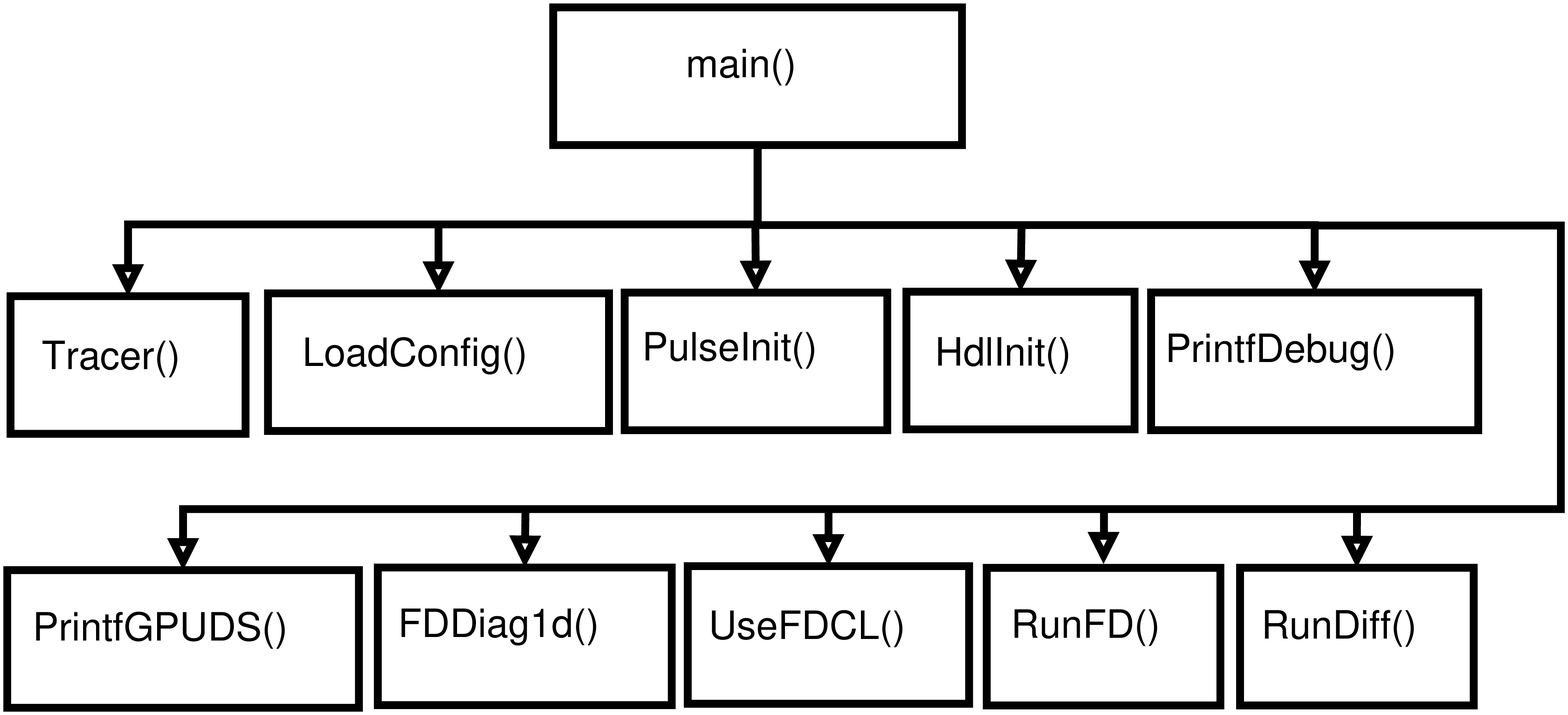}
\caption{The first level of the call graph for the main() C function.  \textit{Tracer}, \textit{PrintfDebug},\textit{ PrintGPUDS} provide simple debugging. \textit{LoadConfig} is the entry point into the \textit{Configuration file} reading. \textit{PulseInit} initialises the Pulse structures. \textit{HdlInit} initialise the OpenCL structures. \textit{FDDiag1d}, and \textit{UseFDCL} initialise the necessary structures, compile the specific OpenCL C code, and allocate the required OpenCL memory objects. \textit{RunFD} is the entry point into running the TDSE propagator and\textit{ RunDiff} is the entry point into the Diffusion equation propagator.}
\label{fig:main}
\end{figure}

There are 5 broad areas within the code;
\begin{enumerate}
\item \textbf{Input}: \textit{ConfigReader.c}, \textit{ConfigReader2.c}
\item \textbf{Output}: \textit{Output.c}
\item \textbf{Pulse}: \textit{PulseInit.c}, \textit{Pulse.c}, \textit{Integrators.c}
\item \textbf{OpenCL}: \textit{clFunc.c}, \textit{clDebug.c}, \textit{clInit.c}
\item \textbf{TDSE}:  \textit{FiniteDiff.c}, \textit{UseFDCL.c}, \textit{Taylor.c}, \textit{Lanczos.c}, \textit{RungeKutta.c}
\end{enumerate}

The \textbf{OpenCL} and \textbf{TDSE} sections rely on the \textit{OpenCL framework}. The \textbf{Input} section is independent of the computing framework used for the propagation, while  the \textbf{Output}, and \textbf{Pulse} sections make use of the OpenCL double precision datatypes for compatibility but are otherwise easily separated from the framework with only minimal effort required.

\subsection{\textbf{Input}}
Initially the \textit{configuration file} (examples are shown later), which is specified as an argument to the program, is loaded into structures in memory. The configuration file provides the necessary information to specify the problem fully. The ability to read the configuration file for \prog comes from the \textit{libconfig} library \cite{libconfig}. Thus the syntax is that listed by the libconfig maintainers. libconfig based configuration files are easy to follow in their structure and also extensible to unanticipated configurations. In order for the configuration file to be read, \prog contains two sections (logically separated and in separate files):
\begin{description}
\item[ConfigReader2.c] provides generic functionality to read configuration files and place the data into the appropriate memory location based on a passed array of structures. Each member of the array represents a specific setting. Each array of structures is terminated with a NULL structure.

\item[ConfigReader.c] is set to pass the specific arrays of structures required by \textit{ConfigReader2.c} for reading settings. This array of structures approach is only used where feasible. It is sometimes more practical for the functions in \textit{ConfigReader.c} to access the libconfig functions directly, such as for the necessary translation of strings to enumerated types which are used internally within \prog. The tedious conversion of converting character arrays to enumerated types is done to facilitate the ease of future comparisons and the ease of using switch statements etc. Performing the necessary conversion in a single location provides clarity.
\end{description}

This configuration file handling is an example of the \textit{reusability} design of the code. The functions of \textit{ConfigReader2.c} provide extensible configuration reading, whilst \textit{ConfigReader.c}, higher in the function tree, has more problem specific and less generic functionality. This then lends itself to modularity and possible code re-use for parts of the code.

\subsection{\textbf{Output}}

\begin{table}
\begin{tabular}{ | l | l | }
\hline
\textbf{File name} & \textbf{Description} \\
\hline
laser.dat & The laser pulse, including the electric field, vector potential and the intensity. \\
output.dat & Time dependent functions which return only a single value per time step. \\
current.dat  & The angle dependent probability current through a fixed radius. \\
radial$t$.dat  & The time dependent radial distribution at a particular time $t$. \\
ProbCurr\_l$l$.dat & The angle dependent probability current density for a particular partial wave value $l$. \\
\hline
\end{tabular}
\caption{Output files after the propagation of the CLTDSE executable.}
\label{tbl:output}
\end{table}

The produced output files are given in Table \ref{tbl:output}. In particular, the file \textit{output.c}, contains the following externally facing (in terms of the API) functions:
\begin{description}
  \item[PrintFD()] outputs the angle-integrated population, as well as the population of the angle integrated partial waves $\abs{f_{l0}(r_j, t)}^2$, at each radius $r_j$. Currently this is only used in the diffusion equation function \textit{clDiffusion}.
  \item[jPrintCurrentFD()] calculates the angle-dependent radial probability current, as given by Eq. \ref{eq:flux} at a particular radius $r_b$  (determined by the configuration file setting IonPos) for each time step, is printed to a file. Graphs such as Fig. \ref{fig:RadCurr} can be produced with the data (in this instance using R).
  \item[PrintHarmonic()] prints the population of the gauge dependent partial waves $\abs{\frac{1}{r}f^{(G)}_{l,0}(r,t)Y_{l0}(\theta,\phi)}^2$ in terms of $\theta$ and $r$. The data is printed into separate files of the form ProbCurr\_l$l$.dat, where $l$ is the angular momentum.
  \item[PrintOut()] is responsible for printing all the time dependent functions which return a single value at each time step to a file. Values are printed in the following order: the time, the ground state population Eq. \ref{eq:p_g}, the ionisation yield estimate Eq. \ref{eq:yield_1}, the excited state estimate Eq. \ref{eq:p_e}, the dipole moment Eq. \ref{eq:Dipole}, the dipole velocity Eq. \ref{eq:DipoleVelocity}, the dipole acceleration Eq. \ref{eq:DipoleAccel}, the ionisation current at a fixed radius Eq. \ref{eq:yield}, the norm Eq. \ref{eq:Norm}, and the difference between unitarity and the actual norm (with the difference being due to numerical limitations and the absorbing potential, if applicable). These values are printed to the file \textit{output.dat}.
  \item[PrintPulse()] is also an output function but unlike the others it is located in Pulse.c. For each time step, the time t, the electric field E(t), the vector potential A(t), and the intensity I(t) are printed out to the specified file.
\end{description}

\subsection{\textbf{Pulse}}
After the configuration file has been loaded, the pulse is initialised:
\begin{description}
\item[PulseInit()] in \textit{PulseInit.c} is the entry point into pulse initialisation. The pulse initialisation includes (a) specifying the gauge of the problem, (b) the form of the pulse(s), (c) conversion of settings to atomic units, if applicable, and (d) the calculation of the length of time required for propagation.
\end{description}

During the simulations, when the electric field $E(t)$ (\textit{PulseE()} in \textit{Pulse.c}) or the velocity gauge vector potential $A(t)$ (\textit{A()} in \textit{Pulse.c}) is required, the pulse is calculated on the CPU within the host code and transferred to the specific compute device.

\subsection{\textbf{OpenCL}}

Next we describe some important functions and structures that are essential for the establishment of the OpenCL framework, and important for queueing kernels during execution. The actual calculations are performed through OpenCL C code. The C code merely deals with management issues such as copying arrays and passing in new pulse values, as well as queuing kernels for execution. All mentioned functions are located in \textit{clFunc.c} unless otherwise noted.

\subsubsection{Initialisation}

\begin{description}

\item[clHandle] is a structure used to store handles and objects returned from OpenCL API function calls. Initialisation functions return relevant handles to relevant devices, queues, contexts etc. These handles are stored within a single structure, designated as type clHandle. This allows the construction of an API where internal layout is irrelevant to those writing functions which make use of the framework. After pulse initialisation, OpenCL is initialized and a \textit{clHandle} structure is allocated for use by calling the function \textit{HdlInit()}.

\item[allocHdl()] follows the specific sequence of function calls to initialise OpenCL which are laid out in the OpenCL specification. Prior to calls to the OpenCL API,\textit{ allocHdl} is called to allocate memory for the new handle, the platform and device Ids. \textit{allocHdl} also initialises the unused pointers and variables within the \textit{clHandle} structure to \textit{NULL} and \textit{zero} respectively. The number of programs that will be compiled must also be specified at this point. A program is defined, in this context, as a set of OpenCL kernels which are executed to perform one task; for example in the current case there are \textit{Taylor}, \textit{Lanczos}, and\textit{ Runge-Kutta} programs. allocHDL relies on the following OpenCL library function calls:

\begin{description}
\item[clGetPlatformIDs()] A platform ID is acquired through the OpenCL function \textit{clGetPlatformIDs()}. In principle this function can be used to select between different implementations of the OpenCL library.

\item[clGetDeviceIDs()] This allows compute devices of a specific type to be selected. The three device types are\\ CL\_DEVICE\_TYPE\_CPU, CL\_DEVICE\_TYPE\_GPU and CL\_DEVICE\_TYPE\_ACCELERATOR. In the present case the latter has not been tested as an option (due to the lack of non-GPU accelerators available). Devices can be further split into sub-devices, although for the present case this is undesirable and so has not been performed.

\item[clCreateContext()] associates the selected devices together into a single context. Later, when kernels are compiled, they will be available for execution on any device within the context. A callback function can be associated with the context to provide error handling. The necessary function pointer is passed to the \textit{clCreateContext()} function.

\item[clCreateCommandQueue()] creates a queue. For every compute device within the context, a separate queue is created. Kernels are added to this queue for execution on the associated compute device. Currently only one compute device at a time is supported in \prog.
\end{description}

\item[UseFDCL()] in \textit{FiniteDiff.c} allocates and initialises the structures necessary for finite difference approach.
\end{description}

\subsubsection{Kernel execution}
\begin{description}
\item[clExecuteSizedKernel()] is a minimal function which calls \textit{KernInit()} to ensure the memory objects for the kernel of interest are initialised, and to also call the OpenCL function \textit{clEnqueueNDRangeKernel()} to enqueue a kernel. Kernel execution is asynchronous; that is, when you enqueue a kernel to be executed it is not necessarily executed before the \textit{clEnqueueNDRangeKernel()} function returns. This ensures that the GPU can be throughput dependent rather than being latency dependent; it can build up a batch of kernels and then pass the data over the PCI bus. During kernel execution, at occasional intervals, the radial grid array is read and processed for output. When the kernels are being executed a number of functions provide information on the dimensions of the problem, while other functions provide the specific place of each instance of the kernel within the problem. The OpenCL function \textit{clEnqueueNDRangeKernel()} is always called with the number of work groups and the number of work items as arguments.

\end{description}
\subsubsection{Work Items}
\begin{description}
\item[SetupGroups()] dicates the work group size. When OpenCL kernels are executed they are not executed in a single thread but simultaneously by multiple ``threads'' known as work items; i.e an instance of a kernel is called a work item. Work items execute instructions in a SIMD fashion with a collection of other work items called a work group. In the host code, an ideal number of work groups and work items must be passed as arguments to \textit{SetupGroups()}. In \textit{SetupGroups()}, through a call to \textit{clReqSizeTInfo()} to get the maximum allowed work group size, the ideal value is modified to take into account the work group size limits. The function \textit{SetupGroups()} sets up a default configuration of work groups and their sizes for the calls to \textit{clExecuteKernel()}. This default configuration can be overlooked by executing kernels through \textit{clExecuteSizedKernel()}.
\end{description}

Within an executing OpenCL kernel:
\begin{description}
 \item[get\_global\_id()] provides a number identifying a particular work item with respect to the other work items. The integer it returns ranges from $0$ to \textit{get\_global\_size()} $- 1$.
 \item[get\_local\_id()] provides the number of the work item with respect to the other work items in it's work group. The integer returned ranges from $0$ to \textit{get\_local\_size()} $- 1$.
\end{description}

A work group is simply a logical grouping of work items that also share a common local memory. Since problems can be multi-dimensional, the parameter of the above functions is an integer indicating the particular dimension. For the parellelisation of the calculation of a new vector only 1 dimension is required, so 0 is used as the argument to refer to that 1 dimension, e.g \textit{get\_global\_id(0)}.

$\v{F}_{l0}(r_j, t)$ is dependent on a value of $l$ and $j$, both of which are integers. The simple approach of calculating $l$ and $j$ can be expressed in 2 very simple lines. $l$ indicates the angular momentum and it is simply calculated by dividing the global id by the number of grid points per $l$ block (\textit{GRID\_SIZE}): \begin{algorithmic}  \STATE $l \gets get\_global\_id(0)/GRID\_SIZE$ \end{algorithmic} Note that in this division we are relying on a property of integer division. In C, division of integers is not like the standard floating point division. Integer division rounds down after dividing two integers, i.e $(int)3/(int)2 = 1$.

If we know the particular angular momentum $l$ and if the value is multiplied by the number of states $l*GRID\_SIZE$, that gives the global id corresponding to the start of the block. If we subtract that value from the global id: \begin{algorithmic} \STATE $j \gets get\_global\_id(0) - l*GRID\_SIZE$ \end{algorithmic} we find the spatial coordinate $j$. If this property of integer division was not used then it is clear that $j \equiv 0$.

Previously, when the kernels were executed the assignment of work was calculated according to the algorithms, as discussed in \cite{OBroinNikolopoulos2012}, which ensured that for each coefficient to be calculated (via a dot product) there is one thread. It also ensured that the threads were logically grouped  into work groups by the part of the array they are calculating. Although the complexity of the method added no tangible computational burden for the case of the basis set, a much simpler approach is used for finite difference because there is no large Hamiltonian stored in memory to read from.

\subsubsection{Compilation and object initialisation}
As well as setting up the work group structure in the host code, the specific kernels that will be used for finite difference must be compiled. Several functions help to wrap up the OpenCL build functions, and to manage the OpenCL memory objects:

\begin{description}
\item[BuildInitKern()] is the highest function in the tree for compilation, it calls the necessary subfunctions which handle compilation (\textit{clBuildFile()}), kernel object allocation and initialisation (\textit{SetupKerns()}).
\item[clBuildFile()] takes the specified OpenCL C source files and any additional strings specified and calls the necessary functions to load and compile them. Programs, and thus kernels, are compiled for every device within a specified OpenCL context. clBuildFile is associated with a specific program with the total number of programs having been decided by \textit{HdlInit()}. \textit{clBuildFile()} relies on the following OpenCL functons:
\begin{description}
 \item[clCreateProgramWithSource()] prepares the source file for compilation and returns a program handle.
 \item[clBuildProgram()] performs the actual compilation associated with a program handle.
\end{description}
\item[clBuildInfo()] provides any errors or warnings returned by the compilation attempt.
\item[SetupKerns()] allocates the structures that hold the kernel memory objects, and the kernel handles as well. The allocation and initialisation of kernel handles is done through the function \textit{AllocKern()} which, in turn, relies on the OpenCL function \textit{clCreateKernel()}. \textit{AllocKernArgs()} allocates memory in the clHandle structure to link to specific memory objects which are then initialised by \textit{SetupObjs()}.
\end{description}

The specific memory objects and forms of the kernels for the various methods varies greatly; the initialisation functions in \textit{clInit.c}: \textit{clBuildLanczos()}, \textit{clBuildTaylor()}, \textit{clBuildRungeKutta()} are distinct, as well as the code for kernel enqueuing for the Taylor expansion (\textit{Taylor.c}), the Lanczos method (\textit{Lanczos.c}) and the Runge-Kutta methods (\textit{RungeKutta.c}). They share the above functions for initialising, compiling and building which have been implemented and the Pulse API.

Currently the code assumes that each angular momentum block of the radial vector contains the same number of coefficients, since this suits the particular use case discussed here. To modify this would only require modifying the two lines of code which correspond to defining $l$ and $j$, and possibly the number of work items enqueued via \textit{clEnqueueNDRangeKernel()}.

\subsection{\textbf{TDSE}}

The actual calculations are performed through OpenCL C code. The C code for each method merely deals with management issues such as copying arrays and changing the pulse values, as well as queuing kernels for execution. For the propagation schemes implemented in the code, the main bottleneck operation is the matrix-vector multiplication. In particular, we have implemented (a) the Runge-Kutta methods, (b) the Taylor series, and (c) the Krylov based Lanczos algorithm methods. The Taylor and Runge-Kutta kernels described here, are the same as those described in \cite{OBroinNikolopoulos2012}, except that the kernels have now been re-purposed for finite difference.

\paragraph{Taylor.c}

The main feature of the Taylor series propagator is its simplicity. Unfortunately as a method, it is not always stable, although it  converges quite well when propagating the hydrogen system. For helium systems, as in \cite{SmythEtAl1998}, the Taylor series was also noted to be very reliable. Considering the propagation of Eq. \ref{eq:tdse_matrix}, we evaluate the vector $\v{F}_{n+1} = \v{F}(t_n + dt)$ by evaluating the following Taylor expansion:
\begin{equation}
\v{F}_{n+1} = \sum_{p=0}^{P} \v{F}^{(p)}_n \qquad
\v{F}^{(p+1)}_n = \frac{-idt}{p}\v{H}\cdot \v{F}^{(p)}_n.
\label{eq:Taylor}
\end{equation}
Since the Taylor expansion of the exponential operator is such a simple expression, it also leads to very simple OpenCL C code. Only one kernel is required, which performs the calculation given by equation \ref{eq:Taylor} and adds it to the solution for $\v{F}(r, t+dt)$.

\paragraph{RungeKutta.c}

The Runge-Kutta methods are of the form:
\begin{equation}
\v{F}_{n+1} = \v{F}_{n} + h \displaystyle \sum \limits_{i=1}^S b_{i} \v{k}_i, \qquad
\v{k}_i = \v{F} \left ( t_{n} + c_i h, \v{F}_{n} + h \displaystyle \sum \limits_{j=1}^{i-1} a_{ij} \v{k}_j \right ).
\label{eq:RKDeriv}
\end{equation}
where $b_i, c_i, $ and $a_{ij}$ are values dependent on the specific method.

The Runge-Kutta algorithm is broken into 3 kernels. One kernel performs the summation on the RHS of \ref{eq:RKDeriv}, while the second kernel performs the derivative calculation. Finally, the third kernel adds the linear combination of multiple derivatives. The pseudocode for each time step is:

\begin{algorithmic}
\STATE $i \gets 0$
\WHILE{$i < P$}
 \STATE $\v{g} \gets \v{f}_{n} + h \sum \limits_{j=1}^{i-1} a_{ij} \v{k}_j$
 \STATE $\v{k}_i \gets \v{f} \left ( t_{n} + c_i h, \v{g} \right )$
 \STATE $i \gets i+1$
\ENDWHILE

\STATE $\v{f}_{n+1} \gets \v{f}_{n} + h \sum \limits_{i=1}^S b_{i} \v{k}_i$
\end{algorithmic}
Since, for each derivative calculation, a vector potential is calculated with a Gaussian quadrature and then added to the previous vector potential, care must be taken to ensure the vector potential is handled correctly. Subtle errors can be introduced by an incorrect sequence. Since we are not concerned about explicitly including the speed of light term $c$, using the substitution $\tilde{A}(t) = A(t) / c$, the calculation performed during the derivative calculation to be approximated is:  $\tilde{A}(t_{n} + c_i h) = \tilde{A}(t_{n}) + \int_{\tau = t_{n}}^{t_{n} + c_i h} d\tau E(\tau)$. Ignoring the $c$ term, after each time step a calculation much like Eq. \ref{eq:VectorStep} has been performed.

\paragraph{Lanczos.c}
Lanczos propagation, here, refers to the application of the Lanczos method to generate a smaller system that can then be propagated with other standard methods. The kernel for the Lanczos method is also more complicated in comparison to the RK and Taylor kernels.

The Lanczos algorithm is broken into 5 kernels. Multiple kernels are required because global memory synchronisation only occurs between all work groups after the execution of a kernel. That is, only local synchronisation, which occurs between members of the same work group, can occur during kernel execution. This restriction is present to easily allow for different execution models with no impact on the results. Different methods of execution include queueing the kernels to execute serially, in parallel, or both, depending on the compute device.

The first kernel, \textit{LanczosLoop1()} performs the necessary power-iteration matrix vector calculation; although in the current context the matrix is not explicitly stored. The kernel also starts the two-kernel process of calculating $\alpha$ through reduction. Two kernels are required since we need to perform a reduction across work groups. In the first function a local reduction is performed for each work group. Then a second reduction is done in \textit{LanczosLoop2()} with each work group to independently calculate $\alpha$.

For a GPU, there is a standard method for performing a reduction. It is a method of performing a dot product that relies on as little communication, and thus synchronisation, as possible. Local memory is used for the reduction where possible because access latency is lower.

Initially items are read from the global array pArr[] into the local Work[] in such a way that if the global memory is larger than the local memory each work item adds the correct items. Abbreviating get\_local\_id(0) by LID, and get\_local\_size(0) with LSZ, and where Len is the length of the array to be added, we have:
\begin{lstlisting}
    int NumTile = Len / LSZ;

    Work[LID] = (LID < Len ? pArr[LID] : 0.0);
    for (int i = 1; i < NumTile; i++)
        Work[LID] += pArr[i*LSZ + LID];

    int Rem = (NumTile + 1) * LSZ + LID;
    if (Rem < Len)
        Work[LID] += pArr[Rem];
\end{lstlisting}

After this, a for loop is entered, where at each step, the number of active work items is reduced by half. This is done by bit shifting the value of an integer $i$ initially set to half the work size down 1 bit each step in a for loop, which is equivalent to division by 2. On each loop, each work item checks if its local Id is less than the loop iterator value $i$. If the local Id is less than the loop iterator, then the work item grabs an array element which is $i$ further down the array from the element in position $LID$, and adds it to the element in position $LID$:

\begin{lstlisting}
    for (unsigned int i = LSZ >> 1; i > 0; i >>= 1)
    {
        barrier(CLK_LOCAL_MEM_FENCE);

        if (LID < i)
            Work[LID] += Work[LID + i];
    }
    barrier(CLK_LOCAL_MEM_FENCE);
\end{lstlisting}
The work items in the work group, some of which are performing no work, must still all reach the local memory synchronisation barrier.

As well as performing the $\alpha$ calculation in \textit{LanczosLoop2()}, we perform the orthogonalisation procedure on the new vector in the Krylov subspace, as well as performing the first step in a reduction to calculate $\beta$.

To complete the Krylov subspace algorithm, in \textit{NormKrylov()} the value $\beta$ is calculated through the parallel normalisation of the new vector by reduction.

The function \textit{Taylor1()} implements a simple parallel algorithm for the Taylor series which uses only one work group, while Taylor2 calculates $\v{f}(t+dt)$ from $\tilde{\v{f}}(t+dt)$ (the expansion from the subspace back to the state space).

\section{Using the Program} \label{sec:User}

\subsection{Installation} \label{sec:Install}
The Installation of OpenCL is hardware and distribution dependent. For standard CPU and AMD GPU usage, the AMD Accelerated Parallel Processing (APP) package should be used (currently on v2.8). Alternatively the Intel SDK can be used for CPU only. The NVIDIA OpenCL driver only works on Nvidia GPUs. OpenCL support is built into the standard Nvidia driver. Nvidia Optimus systems may also require the \textit{Bumblebee} program since the GPU switching technology is currently not implemented for Linux based systems. The installation of Bumblebee is system dependent.

The package \textit{libconfig} should also be installed for the configuration file processing. This package is available for most common GNU/Linux distributions.

The distributed archive contains the source code (\textit{src}), the OpenCL source code (\textit{osrc}), a directory for the binary (\textit{bin}) and a run directory (\textit{run}). \prog should be run from the run directory. The run directory contains a symbolic link to the OpenCL source code directory of the same name, a directory for holding input configuration files \textit{inp}, a directory for holding the generated ground state and for output \textit{out}.

In the \textit{src} directory, the commands
\begin{verbatim}
cltdse_fd/src> make clean ; make cleansrc
cltdse_fd/src> make CLTDSE
\end{verbatim}
remove the object files and most of the temporary files generated by text editors and those files generated by clang during analyse mode, and then compiles the code.

For the makefile,\textit{ gcc.cfg} and \textit{clang.cfg} are provided. CFLAGS dictates the compiler flags used. Two options are given for each compiler, one for debugging, and an option for regular optimised compilation. Uncomment the desired flag (and comment out the other flag when applicable).

Execution of the program is quite simple. The binary is run with the location of a specific configuration file as a parameter, e.g:
\begin{verbatim}
cltdse_fd/run> bin/CLTDSE inp/Diff1.cfg
\end{verbatim}

The configuration file contains all of the relevant information that describes the problem to be simulated.

Each example given is a progression from the previous example to add additional complexity. In the configuration files, a number of settings are specified to handle the location of files for output, the pulse, the size of the system, as well as the number of discrete points represented and settings for the propagator itself.

\begin{figure}[!t]
\centering
\includegraphics[width=300px]{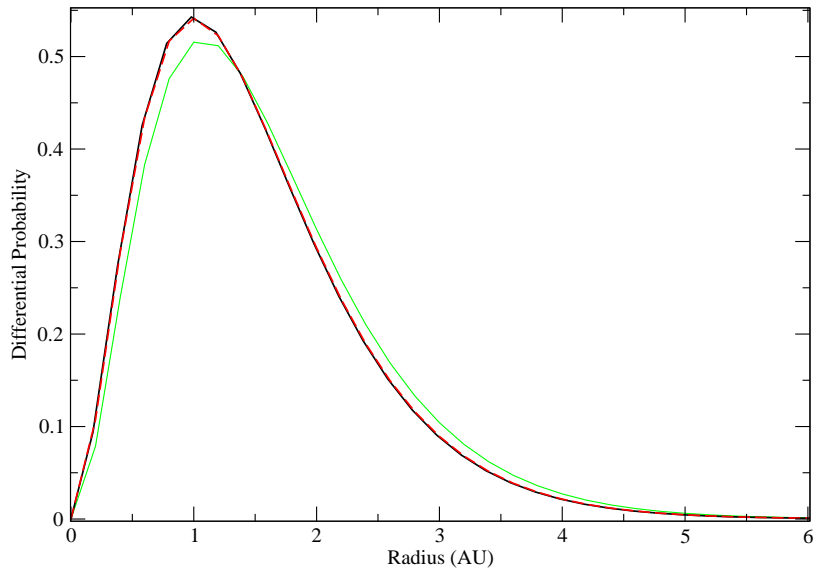}
\caption{The differential probability for radial function for the 1s ground state of hydrogen is calculated with a grid of size $800$ au and $dr = 0.2$ au. The initial calculation without an offset is given by the green line. The system with an offset is given by the black line. The dashed red line indicates the analytical radial probability given as $4r^2 e^{-2r}$ in atomic units. }
\label{fig:1sHydroRad}
\end{figure}

\subsection{Example 1:Example1.cfg}

In the example configuration file \textit{inp/Example1.cfg}, an hydrogen system is subjected to an electric field. This file consists of a number of separate entries called \textit{Groups} that serve to configure the run. The following provides a brief description of the various configuration groups included in this example:
\begin{enumerate}
\item[Group] \textbf{Files}: The various setting included in this group define the location of various files as follows:
\begin{itemize}
\item[Setting] \textit{osrc}:  sets the location of the OpenCL C source code directory
\item[Setting] \textit{Data}: sets the location to store the ground state function
\item[Setting] \textit{Laser}: sets the location to print laser output
\item[Setting] \textit{Plot}: sets the 2d plot of partial waves
\item[Setting] \textit{Radial}: sets the location to print the square of the radial wavefunction,
\item[Setting] \textit{Time}: sets the location to print time dependent outputs excluding the radial wavefunction
\item[Option] \textit{Divisions}: indicates the approximate number of discrete points that will be in the time dependent data set, with a minimum value being the same number of points in the propagation scheme
\end{itemize}

\item[Group] \textbf{Output}:
\begin{itemize}
\item[Setting] \textit{TimeDep}: 1 indicates that time dependent output should be printed to file, while 0 indicates no time dependent output is to be printed.
\end{itemize}

Since this output is currently not performed asynchronously, this can provide a decrease in total runtime when the time dependent outputs are not desired.

\item[Group] \textbf{System}:
\begin{itemize}
\item[Setting] \textit{Type}: set to \textit{``Finite''} for the finite difference simulation
\item[Setting] \textit{Equation}: set to \textit{``TDSE''}  for the TDSE calculations or \textit{``Diff''} for the case of the diffusion equation approach for generating the ground state.
\end{itemize}
\item[Group] \textbf{Matrix}:
\begin{itemize}
\item[Setting] \textit{Gauge}: selects the gauge by setting \textit{``Length''} or \textit{``Velocity''}.
\end{itemize}
\item[Group] \textbf{Propagators}: is a group that provides the settings for the time integration.
\begin{itemize}
\item[Setting] \textit{Device}: can be set to \textit{CPU} or \textit{GPU} depending on whether you wish the simulation to be run on the CPU or GPU
\item[Setting] \textit{WorkItems} indicates the number of work items per work group you wish for the simulation. It is generally recommended that the number of work items have at least $32$ as a divisor due to general GPU structures. In the case of AMD graphics cards 64 is the minimum granularity for executing work items; numbers of work items less than 64 simply result in wasted execution cycles (this is referred to as a ``wavefront'' in AMD terminology)
\item[Setting] \textit{Method}: is set to \textit{``Taylor''}. The other settings are \textit{``RungeKutta''} and \textit{``Lanczos''}. (Since \textit{``Taylor''} is selected in the current example, the group \textit{Taylor} is also present here, for which the\textit{ setting} \textit{Order} is set to the order of the Taylor propagator.
\end{itemize}
\item[Group] \textbf{FD}: The properties of the finite difference grid are specified in this group.
\begin{itemize}
\item[Setting] \textit{NumBlocks}: sets the number of angular momentum blocks
\item[Setting]  \textit{Numdx}: sets the number of grid points per angular momentum block
\item[Setting] \textit{Len}: is an array of length 1, indicating the radial dimension of the box in atomic units. Typically while pulse parameters are specified in standard units, the box dimensions are in atomic units. The system is in a ``box'' of radius $800$ au and 4000 grid points, meaning $dr = 0.2$ au.
\item[Setting] \textit{Start}:  Since the finite difference starting location is offset from zero, this indicates the number of grid points the system is shifted away from $r = 0$. For example, if \textit{Start = 1;} is specified then the initial grid point $r_0 = dx$.
\item[Setting] \textit{Offset}: allows a general offset to be specified to apply to the grid, to offset the grid by a specific amount specified in atomic units. Note that if the offset is beyond a small tweak it will mean that a different integration technique would need to be used which is more sophisticated than Simpson's rule.
\end{itemize}
\end{enumerate}

\begin{figure}[!t]
\centering
\includegraphics[width=300px]{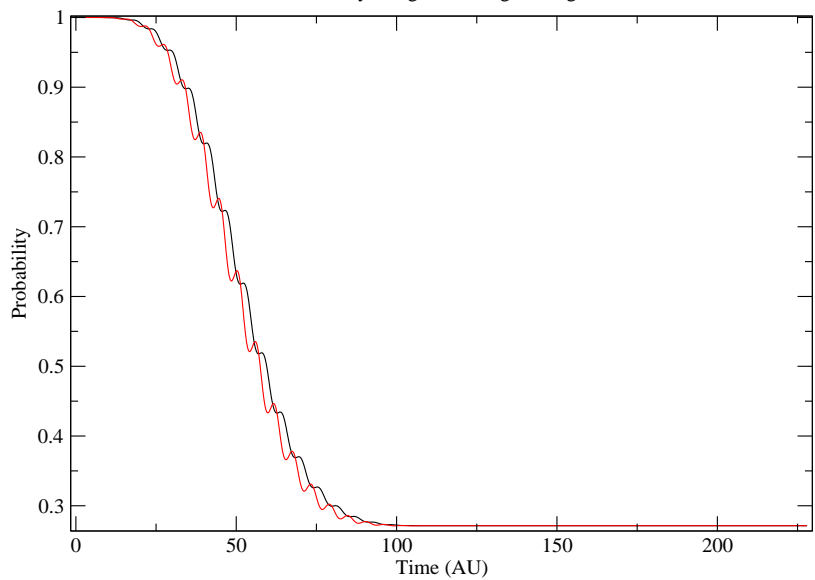}
\caption{The population of the ground state of an hydrogen system as it is under the action of the sine-squared vector potential as given by \ref{fig:VecPot}. Show in red is the length gauge result, while in black the velocity gauge. }
\label{fig:GroundPop}
\end{figure}

After having configured the configuration file for the time-dependent run, we have to prepare a configuration file in order to generate the ground state of the atomic system, in the present case of hydrogen. To this end, the  \textit{inp/Diff1.cfg} configuration file provides the
required settings. For the ground state calculation changing \textit{Equation = ``Diff''} provides enough settings to generate the ground state for the system. Note that the field does not need to be specified at this point.

\begin{verbatim}
cltdse_fd/run> CLtdse inp/Diff1.cfg
\end{verbatim}

Running \prog with the configuration file \textit{inp/Diff1.cfg} will generate and save the radial eigenfunction for the ground state. Without tweaking the inner boundary, the ground state energy is given as approximately $-0.47$ au, and the radial function is as shown in Fig. \ref{fig:1sHydroRad}. It may be noticed that, compared to the analytical ground state, the calculated state is a rather poor match. By setting \textit{Offset} within the group \textit{FD} as is done in \textit{inp/Diff2.cfg}, one can introduce a slight modification. An offset of $-0.02$ au returns a radial function much closer to the analytical function, and also with a closer energy value of approximately $-0.5$ au. From the perspective of an ab-initio calculation of the ground state this tweaking to fit the result is undesirable, but considering that we are interested in the time dependent modification of the states, rather than the ab-initio generation of states, this approach is fairly standard.

At this stage we run the executable with argument the Example1.cfg configuration file. As explained in detail the settings in this file have been selected to support a time-dependent propagation:

\begin{verbatim}
cltdse_fd/run> CLtdse inp/Example1.cfg
\end{verbatim}

\begin{figure}[!t]
\centering
\includegraphics[width=300px]{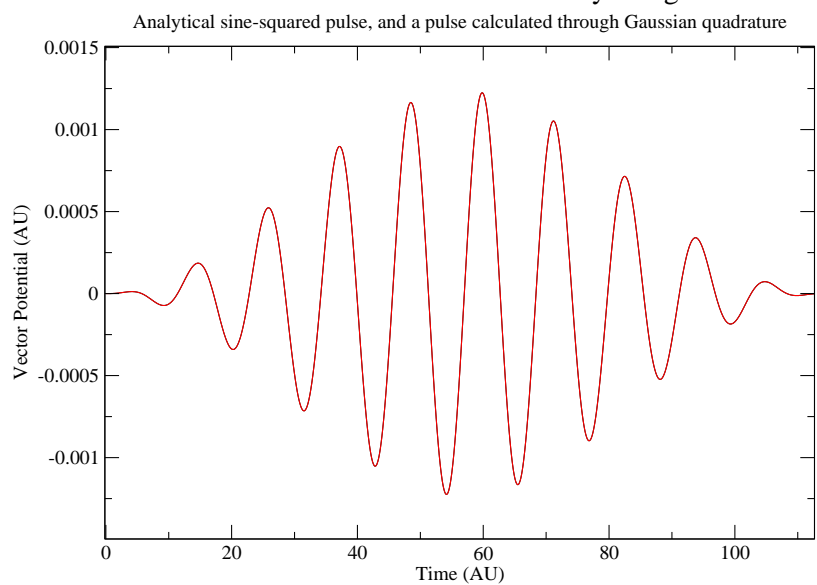}
\caption{The analytical vector potential compared to the vector potential calculated through a 5 point gaussian quadrature. The pulse properties are $1 \times 10^{15}$ $W/cm^2$, 10 $eV$ and 10 cycles. }
\label{fig:VecPot}
\end{figure}

In the particular case of the present example, the vector potential generated from the implemented 5-point gaussian quadrature when compared to the analytical form agrees very well as shown in Figure \ref{fig:VecPot}. The output vector potential does not include the multiplication by the fine structure constant required for atomic units; this multiplication can be performed after the fact.

To generate a pulse of this form, in the group \textit{Field}, an array of groups is defined called \textit{Laser}. Each entry in this array defines the parameters for a particular pulse. In the current example, where the array contains only one group:
\begin{description}
 \item[Units] is set to \textit{``Standard''} to indicate that the numbers given are in standard units for the laser parameters ($eV$, $W/cm^2$, femtoseconds (fs), etc)
 \item[Intensity] is set to \textit{1e15}, which denotes an intensity of $1 \times 10^{15}$ $W/cm^2$
 \item[W] is set to \textit{15.0} for 15 $eV$. The number is written in floating point style to indicate that it is not an integer. Internally the parameters are converted to atomic units (by \textit{UnitsConvert()} in \textit{PulseInit.c}).
 \item[Shape] is \textit{SineA} to indicate that the vector potential should be of sine squared form, thus using Eq. \ref{eq:VecSinSinCosE}.
 \item[SineSqr] is the subgroup that provides further details about $sin^2(x)$ pulse envelopes. The setting used within this group is:
 \begin{description}
 \item[Cycles] which indicates the number of cycles present in the pulse and is set to 10.
 \end{description}
\end{description}

During execution, the pulse generated by the above parameters is printed to the directory specified by \textit{Laser} in the \textit{Files} group. Multiple columns are printed: the time, the electric field, the vector potential, and the intensity.


The evolution of the ground state population, calculated using Eq. \ref{eq:p_g} is shown in Fig. \ref{fig:GroundPop}. While the final ground state population is 0.27145, if the gauge is changed to the length gauge by changing the \textit{Gauge} setting to \textit{``Length''}, the calculated population is 0.27144.

During the calculation the outputted value for the ionisation yield should only be used as an approximate indicator of the fastest components of the yield and thus has limited utility. This is due to the yield calculation requiring the system to continue propagating post pulse so that the continuum components can be spatially separated from the bound states. The yield output is shown in Fig. \ref{fig:Yield}.

\begin{figure}[!t]
\centering
\includegraphics[width=300px]{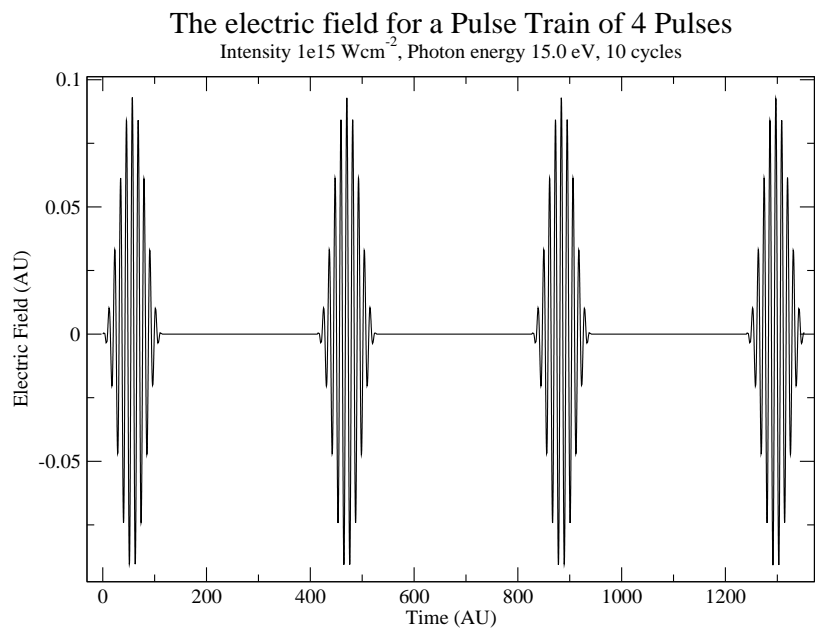}
\caption{The electric field for a pulse train separated by 10 femtoseconds peak to peak. }
\label{fig:PulseTrain}
\end{figure}

\subsection{Other example cases}

By the choice of various settings in the \textit{configuration files} one can investigate different cases. For example in \textit{inp/Example2.cfg} a train of pulses is used, which is shown in figure \ref{fig:PulseTrain}. This is specified by adding the settings \textit{PulseTrain} and \textit{Separation}.
\begin{figure}[!t]
\centering
\includegraphics[width=300px]{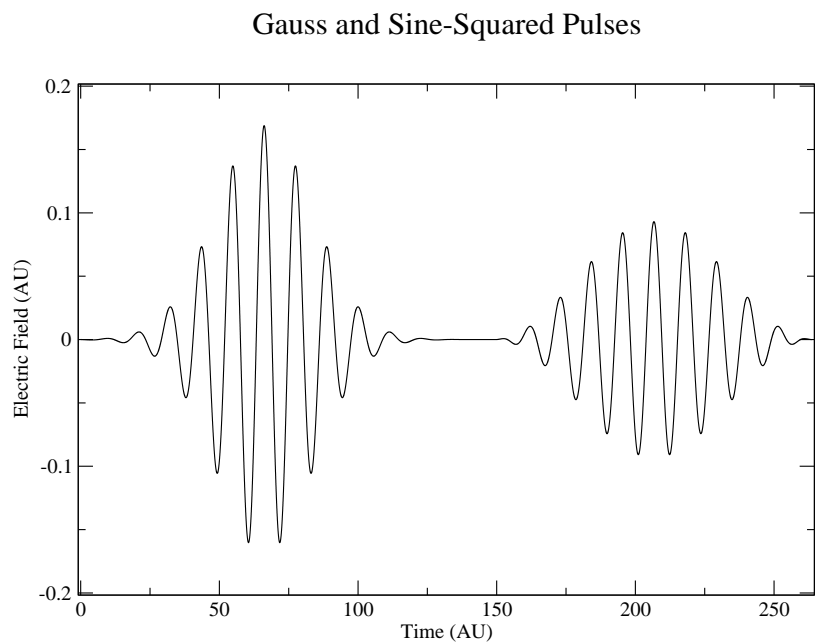}
\caption{The electric field for a Gaussian and a sine-squared pulse. The Gaussian pulse has the same intensity and photon energy as the Sine-Squared pulse, $1 \times 10^{15} W /cm^2$ and $15$ $eV$ respectively, but the full-width half maximum is 2 fs. }
\label{fig:TwoColours}
\end{figure}
To change the pulse given in \textit{inp/Example1.cfg} to a Gaussian form, \textit{Shape} must be changed to \textit{``Gauss''} and the settings \textit{TauD} and \textit{d} must be specified under the subgroup \textit{Gauss} within the \textit{Laser} entry, as shown in \textit{inp/Gauss.cfg}. A Gaussian and a sine-squared pulse can also both be present at the same time as in Fig. \ref{fig:TwoColours}, by simply combining the two separate groups from the  \textit{Laser} entries in \textit{inp/Example1.cfg} and \textit{inp/Gauss.cfg} as is done in \textit{inp/Example3.cfg}. The extra \textit{TimeShift} entry is used to specify the shift of the peak of the second pulse from the $t = 0$ point.

A clear high harmonic spectrum can be produced, as shown in Fig. \ref{fig:HHG}, by increasing the pulse length in \textit{inp/Example1.cfg} to 40 and lowering the intensity. The spectrum is calculated as the Fourier transform of the Dipole moment $D_z(t)$ output (the Fourier transform was carried out using xmgrace). The frequency range can be increased by increasing the number of data points printed. The dipole velocity and dipole acceleration frequency components can, similarly, be plotted.

\section{Benchmarking} \label{sec:Benchmark}
\begin{figure}[!t]
\centering
\includegraphics[width=300px]{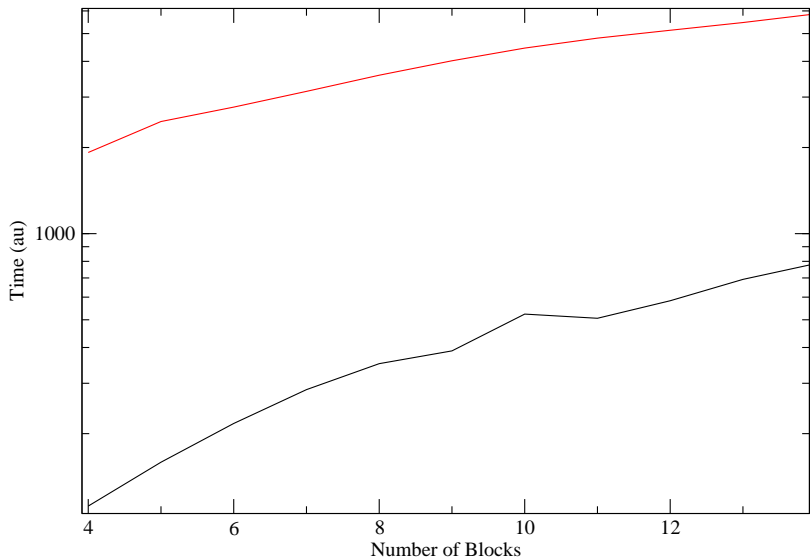}
\caption{ A performance comparison between an 8 hardware thread Intel Xeon with and an AMD 6970 (averaged over three runs each). The average reduction in runtime is 10 times. The y axis is on a log scale. }
\label{fig:Performance}
\end{figure}

It is useful to calculate the relative performance of the CPU to the GPU. To select between GPU and CPU, \textit{Device} in the respective configuration files is changed to \textit{``GPU''} to \textit{``CPU''}. Since real world performance is of most interest, a CPU with several cores and a GPU are compared. For an example benchmark, 2 compute devices are compared: an Intel Xeon E5430 with 4 cores and 8 hardware threads operating at 2.66GHz and an AMD 6970 GPU with 1536 execution units operating at 880MHz. The Xeon E5430 was released at the end of 2007, and the 6970 was released at the end of 2010. The 6970 GPU runs the simulation approximately 10 times faster than the Xeon E5430 as shown in Figure \ref{fig:Performance} on average The exact improvement in execution time is dependent on the system, the more complex the system the better the performance would be expected to be.

While benchmarking, we are interested in the total execution time of the program with a specific configuration. Example configuration files, named \textit{benchmark/Bench4.cfg} to \textit{benchmark/Bench15.cfg} contain configurations for systems with 4 to 15 angular momenta present. The same stepsize is chosen in each system so that the work load scales linearly with $l$. It is expected that the CPU is at a disadvantage since it must execute the simulation, perform the standard operations of the host operating system, and also share time slices with other programs. This disadvantage should become less of an issue as the number of cores, and thus the computing power, is increased.

\section{Modifying the code} \label{sec:Mod}
For each modification, if a new setting is added, you must add support in the configuration file, where the specific function is dependent on what group you are modifying. New configuration settings will require the addition of code to \textit{ConfigReader.c} so as to support the reading of the setting from the configuration file. If the new setting is to be read, one may need to add a new variable to the configuration structures in \textit{ConfigReader.h}. Reading the manual for libconfig is advised, or alternatively one can use the existing code as a template.

To add a new pulse type, three files must be changed. Firstly, the new Pulse must be added to \textit{Pulse.c} with the same function declaration, and with the function prototype added to \textit{Pulse.h}. The prototype for the pulses must match: \textit{PulsePrec PulseName(RPrec, void *)}. If the pulse requires initialisation these should be handled in \textit{PulseInit.c} using the existing pulses as a template.

Changing the structure of the Hamiltonian in the OpenCL C code in \textit{FD.cl} does not necessarily require any change to the configuration options. Since one element of the solution vector is handled by one work item, changes can be made to the matrix vector calculation relatively easily with little impact outside that function. The function \textit{MatVec()} performs the matrix vector calculation $i\dot{\Phi(r, t)}= H \Phi(r, t)$. To modify the static potential, the function \textit{Vm()} is modified.

If any new memory objects are added to a specific propagator, tit is allocated by adding a line into the structure \textit{Objs} in \textit{clBuildTaylor()}, \textit{clBuildRungeKutta()} or \textit{clBuildLanczos()}. Depending on the circumstances, changes to \textit{ReadyMethod()} may also be required. These functions are located in \textit{clInit.c}. If one intends to read or write data to the new memory objects during the time propagation, then changes to \textit{Taylor.c}, \textit{RungeKutta.c} and \textit{Lanczos.c} may be required. A new propagator can be added by duplicating and then tailoring the host code of an existing propagator. New time dependent outputs can be added by modifying \textit{PrintOut()} in \textit{Output.c}.

\section{Conclusions}
In this paper we have presented, \prog, an OpenCL based program for solving the TDSE using the grid-based finite difference approach. We have included an overview of the underlying theory used to describe and solve the hydrogen system, as well as providing information on the electric field and the vector potential. We have given a short description of the equations used within the code for the calculation of observables and other quantities. Instructions have been included on how to install and use the code. A discussion of the structure of the code has also been provided, where we have elaborated on the specifics of the implementation of the Taylor, Runge-Kutta, and Lanczos methods for OpenCL. We have also provided a benchmark to provide a flavour of the performance improvement obtained through GPGPU programming, achieving a runtime reduction of about 10 times when comparing parallel execution on a CPU and a GPU. GPU acceleration opens up the possibility to study problems that are highly computationally demanding, but can be solved in a reasonable timeframe. \prog can help with this goal.

\paragraph{Acknowledgements}
This work has been funded under a Seventh Framework Programme (FP7) project HPCAMO/256601, and with support from the Irish Research Council 'New Foundations' 2012 scheme. LAAN is supported by the Science Foundation Ireland (SFI) Stokes Lectureship Program, and Cost actions MP1203 and CM1204.

\newpage
\appendix

\section{The configuration file settings} \label{sec:Conf}
In the configuration file there are a number of settings to specify. These settings are expressed in terms of\textit{ Groups}
and the associated \textit{Options}. If an \textit{Options} subgroup is not in use it will not be accessed.

\paragraph{Group Files}
It is desirable for the location of input and output files to be fully customizable. As such the Files group allows locations for key input/output operations to be specified.

The Standard settings are: \\
\begin{tabular}{|l |l |p{6.5cm}|}
\hline
  \textbf{Option} & \textbf{Values} & \textbf{Notes} \\
\hline
  osrc & Location string & The location of the OpenCL C files \\
\hline
  Data & Location string & The location of the input data (generated by diffusion equation)\\
\hline
  Laser & Location string & The output location for the laser \\
\hline
  Plot & Location string & The output location for 2D plots \\
\hline
  Radial & Location string & The output location for 1D plots of the radial probability \\
\hline
  Time & Location string & The output location for time dependent quantities \\
\hline
  Divisions & Integer & The approximate number of time dependent outputs that should be made \\
\hline
\end{tabular} \\

\paragraph{Group System}
The Standard setting is:\\
\begin{tabular}{|l |l |p{6.5cm}|}
\hline
  \textbf{Option} & \textbf{Values} & \textbf{Notes} \\
\hline
  Equation & ``TDSE'', ``Diff'' & Whether the equation is the TDSE or the diffusion equation (Ground state calculator) \\
\hline
\end{tabular}

\paragraph{Group Matrix}
The Standard setting is:\\
\begin{tabular}{|l |l |p{6.5cm}|}
\hline
  \textbf{Option} & \textbf{Values} & \textbf{Notes} \\
\hline
  Gauge & ``Length'', ``Velocity'' & The gauge for the system \\
\hline
\end{tabular}

\paragraph{Group Pulse}
The Standard setting is:\\
\begin{tabular}{|l |l |p{6.5cm}|}
\hline
  \textbf{Option} & \textbf{Values} & \textbf{Notes} \\
\hline
  Laser & Array of Groups & Each group specifies a separate laser pulse, or train of pulses.\\
\hline
\end{tabular}

The subgroup Laser has the further settings:\\
\begin{tabular}{|l |l |p{6.5cm}|}
 \hline
  \textbf{Option} & \textbf{Values} & \textbf{Notes} \\
\hline
Units & ``Atomic'', ``Standard'' & The form of the laser units, atomic units or standard $Wcm^{-2}$, eV, fs units. \\
\hline
Intensity & Floating point & Intensity in the respective units \\
\hline
W & Floating point & Photon energy in the respective units \\
\hline
Shape & ``SineSqr'', ``Gauss'', ``SineA'', ``FileE'' & The shape of the pulse \\
\hline
SineSqr & Group & This group is required if ``SineSqr'' or ``SineA'' is selected for Shape \\
\hline
Gauss & Group & This group is required if ``Gauss'' is selected for Shape \\
\hline
\end{tabular}\\

The subgroup SineSqr has the further settings:\\
\begin{tabular}{|l |l |p{6.5cm}|}
 \hline
  \textbf{Option} & \textbf{Values} & \textbf{Notes} \\
\hline
 FineShape & ``cos'' , ``sin'' & The fine shape of the pulse is cosine or sine \\
 \hline
 Cycles & Integer & The number of cycles in the pulse \\
\hline
\end{tabular}\\
The subgroup Gauss has the further settings:\\
\begin{tabular}{|l |l |p{6.5cm}|}
 \hline
  \textbf{Option} & \textbf{Values} & \textbf{Notes} \\
\hline
 TauD & Floating point & The full width half maximum of the pulse. \\
 \hline
 d & Floating point & The strength of pulse chirping \\
\hline
\end{tabular}\\


\paragraph{ Group Propagator}
The Standard settings are:\\
\begin{tabular}{|l |l |p{6.5cm}|}
\hline
  \textbf{Option} & \textbf{Values} & \textbf{Notes} \\
\hline
  Device & ``CPU'' or ``GPU'' & CPU or GPU execution \\
\hline
  Method & ``Taylor'', ``Runge-Kutta'', ``Lanczos'' & Chooses between the different propagators \\
\hline
  StepSize &  Floating point &  Indicates the initial step size in atomic units \\
\hline
  WorkItems & Integer & The number of work items in a work group \\
\hline
  Taylor & Group & This group is required if the Taylor method is specified \\
\hline
  RungeKutta & Group & This group is required if the Runge-Kutta method is specified \\
\hline
  Lanczos & Group & This group is required if the Lanczos method is specified \\
\hline
\end{tabular} \\

The subgroup Taylor has the further settings:\\
\begin{tabular}{|l |l |p{6.5cm}|}
 \hline
  \textbf{Option} & \textbf{Values} & \textbf{Notes} \\
\hline
  Order & Integer & The order of the Taylor propagator \\
\hline
\end{tabular}\\

The subgroup RungeKutta has the further settings:\\
\begin{tabular}{|l |p{6cm} |p{6.5cm}|}
 \hline
  \textbf{Option} & \textbf{Values} & \textbf{Notes} \\
\hline
  Method & ``RKF", ``CashKarp", ``ClassicRK4", ``Euler", ``DormandPrince", ``Merson" & These settings represent a variety of standard Runge-Kutta methods.\\
\hline
\end{tabular}\\

The subgroup Lanczos has the further setting:\\
\begin{tabular}{|l |l |p{6.5cm}|}
 \hline
  \textbf{Option} & \textbf{Values} & \textbf{Notes} \\
\hline
  Dim & Integer & The dimension of the Krylov subspace \\
\hline
\end{tabular}
\newpage

\section{Lanczos methods} \label{sec:Lanczos}

The Lanczos method, assuming perfect precision, is given by\cite{MohankumarAuerbach2006, ANLADemmel1997}:
\begin{algorithmic}
  \STATE $Q_1 \gets \frac{F(t)}{||F(t)||}$
  \STATE $\beta \gets 0$
  \STATE $Q_0 \gets 0$
  \FOR{$p=0$ to $P$}
   \STATE $Q_p \gets \v{H} Q_{p-1}$
   \STATE $\alpha_p \gets Q_{p-1}^T Q_p$
   \STATE $Q_p \gets F - \alpha_p Q_{p-1} - \beta Q{_p-2}$
   \STATE $\beta \gets ||f_p||$
   \IF{$\beta = 0$}
    \STATE quit
   \ENDIF
   \STATE $Q_{p+1} \gets \frac{Q_{p+1}}{\beta}$

   \STATE $\tilde{H}_{p, p+1} = \tilde{H}_{p, p+1} = \beta$
   \STATE $\tilde{H}_{p, p} = \alpha$
  \ENDFOR
\end{algorithmic}

\section{Runge-Kutta} \label{sec:RungeKutta}
A number of common Runge-Kutta integration methods have been implemented: Euler, Classic RK4, Runge-Kutta Fehlberg (\textit{RKF 4(5)}), Dormand-Prince, Cash Karp, and Merson methods. For the classic RK4 there is only one solution, but embedded pair methods provide two solutions, one of higher order $\v{F}^{(2)}$, and another solution of lower order $\v{F}^{(1)}$:
\[
\v{k}_i = \v{f} \left ( t_{n} + c_i h, \v{F}_{n} + h \displaystyle \sum \limits_{j=1}^{i-1} a_{ij} \v{k}_j \right )
\]

\begin{eqnarray}
\v{F}^{(1)}_{n+1} = \v{F}_{n} + h \displaystyle \sum \limits_{i=1}^S b^{(1)}_{i} \v{k}_i \\
\v{F}^{(2)}_{n+1} = \v{F}_{n} + h \displaystyle \sum \limits_{i=1}^S b^{(2)}_{i} \v{k}_i
\end{eqnarray}

As discussed at \cite{NumericalRecipes2007}, the difference between the lower order and higher order solutions,
\[
 \v{\Delta} = \v{F}^{(2)}_{n+1} - \v{F}^{(1)}_{n+1}
\]
provides an estimate of the average deviation from truncation over the entire solution vector,
\[
  \sigma_n \approx \sqrt{ \frac{1}{S} \abs{\v{\Delta}}^2 } = \sqrt{ \frac{1}{S} \sum_{i=1}^{S} \Delta_{i}^2 }
\]
This local truncation deviation between the two methods is used to approximate the truncation between the approximate solution and the real solution, and can be used to extrapolate the value of the global truncation error. This could be used for dynamical step size support, but it is currently not implemented. At present, the sum of the local truncation error is used passively to provide the global truncation error:
\[
\sigma = \sum_{n} \sigma_n
\]




\bibliographystyle{ieeetr}

\bibliography{BibTeXMasterRecord}

\begin{thebibliography}{10}

\bibitem{Posthumus2004}
J.~H. Posthumus, ``{The dynamics of small molecules in intense laser fields},''
  {\em Reports on Progress in Physics}, vol.~67, no.~5, p.~623, 2004.

\bibitem{WabnitzEtAl2005}
H.~Wabnitz, A.~R.~B. de~Castro, P.~G\"urtler, T.~Laarmann, W.~Laasch,
  J.~Schulz, and T.~M\"oller, ``Multiple ionization of rare gas atoms
  irradiated with intense vuv radiation,'' {\em Phys. Rev. Lett.}, vol.~94,
  p.~023001, Jan 2005.

\bibitem{DundasEtAl2000}
D.~Dundas, J.~McCann, J.~S. Parker, and K.~T. Taylor, ``{Ionization dynamics of
  laser-driven H 2 +},'' {\em Journal of Physics B: Atomic, Molecular and
  Optical Physics}, vol.~33, no.~17, p.~3261, 2000.

\bibitem{PengEtAl2003}
L.-Y. Peng, D.~Dundas, J.~McCann, K.~Taylor, and I.~Williams, ``{Dynamic
  tunnelling ionization of $H_2^{+}$ in intense fields},'' {\em Journal of
  Physics B: Atomic, Molecular and Optical Physics}, vol.~36, no.~18, p.~L295,
  2003.

\bibitem{BarmakiEtAl2003}
S.~Barmaki, S.~Laulan, H.~Bachau, and M.~Ghalim, ``{The ionization of
  one-electron diatomic molecules in strong and short laser fields},'' {\em
  Journal of Physics B: Atomic, Molecular and Optical Physics}, vol.~36, no.~5,
  p.~817, 2003.

\bibitem{AwasthiEtAl2005}
M.~Awasthi, Y.~V. Vanne, and A.~Saenz, ``{Non-perturbative solution of the
  time-dependent Schrödinger equation describing $H_2$ in intense short laser
  pulses},'' {\em Journal of Physics B: Atomic, Molecular and Optical Physics},
  vol.~38, no.~22, p.~3973, 2005.

\bibitem{PalaciosEtAl2005}
A.~Palacios, S.~Barmaki, H.~Bachau, and F.~Mart\'in, ``{Two-photon ionization
  of} $h_2^{+}$ {by short laser pulses},'' {\em Phys. Rev. A}, vol.~71,
  p.~063405, Jun 2005.

\bibitem{PalaciosEtAl2006}
A.~Palacios, H.~Bachau, and F.~Mart\'in, ``{Enhancement and Control of} $h_{2}$
  {Dissociative Ionization by Femtosecond VUV Laser Pulses},'' {\em Phys. Rev.
  Lett.}, vol.~96, p.~143001, Apr 2006.

\bibitem{Sanz-VicarioEtAl2006}
J.~L. Sanz-Vicario, H.~Bachau, , and F.~Mart\'in, ``{Time-dependent theoretical
  description of molecular autoionization produced by femtosecond XUV laser
  pulses},'' {\em Phys. Rev. A}, vol.~73, p.~033410, Mar 2006.

\bibitem{CaillatEtAl2005}
J.~Caillat, J.~Zanghellini, M.~Kitzler, O.~Koch, W.~Kreuzer, and A.~Scrinzi,
  ``{Correlated multielectron systems in strong laser fields: A
  multiconfiguration time-dependent Hartree-Fock approach},'' {\em Phys. Rev.
  A}, vol.~71, p.~{012712}, 2005.

\bibitem{Kulander1987}
K.~C. Kulander, ``{Time-dependent Hartree-Fock theory of multiphoton
  ionization: Helium},'' {\em Phys. Rev. A}, vol.~36, pp.~2726--2738, Sep 1987.

\bibitem{GiovanniEtAl2002}
L.~R. Giovanni~Onida and A.~Rubio, ``{Electronic excitations:
  density-functional versus many-body Green's-function approaches},'' {\em Rev.
  Mod. Phys.}, vol.~74, pp.~601--659, Jun 2002.

\bibitem{OBroinNikolopoulos2012}
C.~{\'O}. Broin and L.~Nikolopoulos, ``{An OpenCL implementation for the
  solution of the time-dependent Schr\"odinger equation on GPUs and CPUs},''
  {\em {Computer Physics Communications}}, vol.~{183}, no.~{10}, pp.~{2071 --
  2080}, {2012}.

\bibitem{GuanEtAl2009}
X.~Guan, C.~Noble, O.~Zatsarinny, K.~Bartschat, and B.~Schneider, ``{ALTDSE: An
  Arnoldi-Lanczos program to solve the time-dependent Schr\"odinger
  equation},'' {\em {Computer Physics Communications}}, vol.~{180}, no.~{12},
  pp.~{2401--2409}, {2009}.

\bibitem{Muller1999}
H.~Muller, ``{An Efficient Propagation Scheme for the Time-Dependent
  Schr\"odinger Equation in the Velocity Gauge},'' {\em Laser Physics}, vol.~9,
  pp.~138--148, 1999.

\bibitem{KulanderEtAl1992}
K.~C. Kulander, K.~J. Schafer, and J.~L. Krause, ``{Time-dependent studies of
  multiphoton processes},'' {\em {Advances in atomic, molecular, and optical
  physics}}, vol.~{Supplement 1}, pp.~{247--300}, {1992}.

\bibitem{DundasEtAlEurPhysJ2003}
D.~Dundas, K.~Meharg, J.~McCann, and K.~Taylor, ``{{Dissociative ionization of
  molecules in intense laser fields}},'' {\em {The European Physical Journal D
  - Atomic, Molecular, Optical and Plasma Physics}}, vol.~{26}, pp.~{51--57},
  {2003}.
\newblock {10.1140/epjd/e2003-00082-0}.

\bibitem{Moore2EtAl2011}
L.~Moore, M.~Lysaght, L.~Nikolopoulos, J.~Parker, H.~van~der Hart, and
  K.~Taylor, ``{The RMT method for many-electron atomic systems in intense
  short-pulse laser light},'' {\em {Journal of Modern Optics}}, vol.~{58},
  no.~{13}, pp.~{1132--1140}, {2011}.

\bibitem{NikolopoulosEtAl2008}
L.~Nikolopoulos, J.~Parker, and K.~Taylor, ``{Combined R-matrix eigenstate
  basis set and finite-difference propagation method for the time-dependent
  Schr\"odinger equation: The one-electron case},'' {\em {Phys. Rev. A}},
  vol.~{78}, p.~{063420}, {Dec} {2008}.

\bibitem{Peng2006}
L.-Y. Peng and A.~F. Starace, ``{Application of Coulomb wave function discrete
  variable representation to atomic systems in strong laser fields},'' {\em
  {The Journal of Chemical Physics}}, vol.~125, no.~15, pp.~154311--154311,
  2006.

\bibitem{BauerKoval2006}
D.~Bauer and P.~Koval, ``{Qprop: A Schr\"odinger-solver for intense laser-atom
  interaction},'' {\em Computer Physics Communications}, vol.~174, no.~5,
  pp.~396--421, 2006.

\bibitem{Birkeland2009}
{Tore Birkeland}, {\em {PyProp - A Python Framework for Propagating the Time
  Dependent Schr\"odinger Equation}}.
\newblock PhD thesis, {University of Bergen}, Dec 2009.

\bibitem{SmythEtAl1998}
E.~S. Smyth, J.~S. Parker, and K.~Taylor, ``{Numerical integration of the
  time-dependent Schr\"odinger equation for laser-driven helium},'' {\em
  {Computer Physics Communications}}, vol.~{114}, pp.~{1--14}, {April} {1998}.

\bibitem{Kobus2013}
J.~Kobus, ``{A finite difference Hartree-Fock program for atoms and diatomic
  molecules},'' {\em Computer Physics Communications}, vol.~184, no.~3,
  pp.~799--811, 2013.

\bibitem{QnDynCUDA}
T.~Dziubak and J.~Matulewski, ``An object-oriented implementation of a solver
  of the time-dependent schr\"odinger equation using the {CUDA} technology,''
  {\em Computer Physics Communications}, vol.~183, no.~3, pp.~800--812, 2012.

\bibitem{IEEE754}
``{IEEE Std 754-2008 (Revision of IEEE Std 754-1985) - IEEE Standard for
  Floating-Point Arithmetic},'' 2008.

\bibitem{BachauEtAlRepProgInPhys2001}
H.~Bachau, E.~Cormier, P.~Decleva, J.~Hansen, and F.~Mart{\'i}n,
  ``{{Applications of B -splines in atomic and molecular physics}},'' {\em
  {Reports on Progress in Physics}}, vol.~{64}, no.~{12}, p.~{1815}, {2001}.

\bibitem{NikolopoulosEtAl2007}
L.~A.~A. Nikolopoulos, T.~K. Kjeldsen, and L.~B. Madsen, ``Three-dimensional
  time-dependent hartree-fock approach for arbitrarily oriented molecular
  hydrogen in strong electromagnetic fields,'' {\em Phys. Rev. A}, vol.~76,
  p.~033402, Sep 2007.

\bibitem{BransdenJoachain2003}
B.~Bransden and C.~Joachain, {\em {Physics of Atoms and Molecules}}.
\newblock {Pearson Education Limited}, 2nd~ed., {2003}.

\bibitem{FlocardEtAl1978}
H.~Flocard, S.~Koonin, and M.~Weiss, ``Three-dimensional time-dependent
  {H}artree-{F}ock calculations: Application to $^{16}\mathrm{O}$ +
  $^{16}\mathrm{O}$ collisions,'' {\em {Phys. Rev. C}}, vol.~{17},
  pp.~{1682--1699}, {May} {1978}.

\bibitem{BandraukEtAl2009}
A.~Bandrauk, S.~Chelkowski, D.~J. Diestler, J.~Manz, and K.~J. Yuan, ``Quantum
  simulation of high-order harmonic spectra of the hydrogen atom,'' {\em
  Physical Review A}, vol.~79, no.~2, p.~023403, 2009.

\bibitem{NikolopoulosEtAl2011}
T.~K. L.~A. A.~Nikolopoulos and J.~Costello, ``{Theory of ac Stark splitting in
  core-resonant Auger decay in strong x-ray fields},'' {\em {Phys. Rev. A}},
  vol.~{84}, p.~{063419}, {Dec} {2011}.

\bibitem{libconfig}
M.~Lindner, ``libconfig--c/c++ configuration file library.''

\bibitem{MohankumarAuerbach2006}
N.~Mohankumar and S.~M. Auerbach, ``{On time-step bounds in unitary quantum
  evolution using the Lanczos method},'' {\em {Computer Physics
  Communications}}, vol.~{175}, no.~{7}, pp.~{473 -- 481}, {2006}.

\bibitem{ANLADemmel1997}
J.~W. Demmel, {\em {Applied numerical linear algebra}}.
\newblock {Society for Industrial Mathematics}, {1997}.

\bibitem{NumericalRecipes2007}
W.~Press, S.~Teukolsky, W.~Vetterling, and B.~Flannery, {\em {{Numerical
  Recipes 3rd Edition: The Art of Scientific Computing}}}.
\newblock {Cambridge University Press}, {3}~ed., Aug. {2007}.

\end{thebibliography}







\end{document}